\documentclass[aps,prc,twocolumn,groupedaddress,floatfix]{revtex4}
\usepackage{amsmath} \usepackage{epsfig} \usepackage{graphics}

\begin{document}

\title{Effective Interactions for the Three-Body Problem}

\author{T. C. Luu} \email[]{tluu@lanl.gov}
\affiliation{Los Alamos National Laboratory, MS-227, Los Alamos,
  New Mexico, 87545}
\author{S. Bogner}
\email[]{bogner@phys.washington.edu} 
\author{W. C. Haxton}
\email[]{haxton@emmy2.phys.washington.edu} \affiliation{Institute for
Nuclear Theory, University of Washington, Box 351550, Seattle, WA
98195} \author{P. Navr$\acute{a}$til } \email[]{navratil1@llnl.gov}
\affiliation{Lawrence Livermore National Laboratory, L-414, P.O. Box
808, Livermore, CA 94551}

\date{\today}

\begin{abstract}
  The three-body energy-dependent effective interaction given by the
  Bloch-Horowitz (BH) equation is evaluated for various shell-model
  oscillator spaces.  The results are applied to the test case of the
  three-body problem ($^3$H and $^3$He), where it is shown that the
  interaction reproduces the exact binding energy, regardless of the
  parameterization (number of oscillator quanta or value of the
  oscillator parameter $b$) of the low-energy included space.  We
  demonstrate a non-perturbative technique for summing the
  excluded-space three-body ladder diagrams, but also show that
  accurate results can be obtained perturbatively by iterating the
  two-body ladders.  We examine the evolution of the effective
  two-body and induced three-body terms as $b$ and the size of the
  included space $\Lambda$ are varied, including the case of a single
  included shell, $\Lambda\hbar\omega=0\hbar\omega$.  For typical
  ranges of $b$, the induced effective three-body interaction,
  essential for giving the exact three-body binding, is found to
  contribute $\sim$10\% to the binding energy.
\end{abstract}


\maketitle

\section{Introduction\label{introduction}}

Techniques in popular use in the nuclear three-body problem include
the Faddeev\cite{Faddeev:1961su}, Green's Function Monte Carlo (GFMC),
and correlated hyper-spherical harmonics expansion methods (see, for
instance, refs.\cite{Carlson:1998qn,Pudliner:1997ck,Kievsky:1994bg}).
These methods have been used with realistic phenomenological
potentials (\emph{e.g.} Av18\cite{Wiringa:1995wb},,
CD-Bonn\cite{Machleidt:1996km}, etc. . .) with strong short-range
interactions, yielding binding energies accurate to within
$0.1$\%\cite{Nogga:2002qp}.

All of these methods treat the bare $NN$ interactions in Hilbert
spaces that are effectively infinite.  Because the complexity of the
Hilbert spaces grows very rapidly with nucleon number A, the extension
of such methods to heavier systems becomes increasingly difficult.
This motivates another approach: Solving A-body problems in more
tractable finite ``included spaces,'' while accounting for the missing
physics of the ``excluded space'' through an effective interaction.
For shell-model-inspired effective theories that define included
spaces in terms of harmonic oscillator (HO) bases, the missing physics
includes both high-momentum and long-wavelength interactions.  The
former are missing because the included space contains only low-energy
oscillator shells, while the latter are connected with the finite
extent of the basis states, which are overconfined in an HO potential.

Shell-model calculations often employ two-body effective interactions,
but these are generally determined phenomenologically.  In recent
years there has been growing success in calculating these effective
interactions directly from the underlying bare interaction in a precise
and systematic way (\emph{e.g.}
refs.\cite{Navratil:1998qb,Navratil:1999pw,Haxton:1999vg,Haxton:2002kb}
and references within).  The purpose of this paper is to extend upon
this work by calculating the more complicated $H_{eff}$ that could be
used in a shell-model-inspired effective theory, through three-body
order in the excluded space..  The inclusion of three-body
contributions is generally believed essential in building in the
density dependence crucial to saturation.

The exact effective interaction for the A-body problem is an A-body
operator.  The calculation of this interaction is clearly as difficult
as solving the original A-body problem in an infinite Hilbert space.
However, there are reasons to hope that such an A-body calculation
might be unnecessary.  This hope depends on the plausible notion that
the effective interaction might converge as an expansion in the number
of nucleons interacting in the excluded space.  Were such an expansion
to converge at the three- or four-body cluster level, then standard
three- or four-body methods could be used to treat the excluded space.
These more complicated but still tractable effective interactions
could then be diagonalized in the shell-model-inspired effective
theories to produce accurate binding energies and wave functions.

There are a couple of arguments that such a scheme might converge.
One is the success of the shell model when phenomenological two-body
effective interactions are used.  This suggests that a good part of
the excluded-space physics is two-body.  Another is the qualitative
argument that short-range clustering is increasingly unlikely as the
size of the nucleon cluster increases: such spacial clustering is
necessary for strong, multi-particle, high-momentum interactions.
Finally, we know that the Pauli principle forbids short-range s-wave
interactions above A=4.

Additional support for this notion comes from recent large-basis
\emph{ab initio} no-core shell-model (NCSM)
calculations\cite{Navratil:1998qb,Navratil:1999pw,Navratil:2003ef}.
These calculations use a large but finite HO Hilbert space.  A
Lee-Suzuki (LS) similarity transformation\cite{Suzuki:1980} coupled
with a folded-diagram sum\cite{Towner:1977} is performed on the bare
NN interaction to yield an energy-independent effective interaction.
Early calculations utilized only the effective two-body interaction as
derived from these transformations.  Hence calculations in large
Hilbert spaces, or included spaces, were usually needed to ensure
convergence\footnote{The rate of convergence also depended on the
  oscillator frequency $\omega$ which, if judiciously chosen, could
  improve convergence greatly.} of the binding energy (\emph{i.e.} the
included space had to be large enough such that the contributions to
the binding energy from the effective three- and higher-body terms
were small).  Recent calculations that now include induced three-body
interactions derived from the LS procedure show significant
improvement in the convergence of binding energies for the alpha
particle and some $p$-shell nuclei\cite{Navratil:1998mr,Navratil:2003ef}.

An alternative approach to calculating the induced three-body
effective interaction was developed in Ref.\cite{Haxton:1999vg} using an
excitation-energy-\emph{dependent} three-body interaction based on the
self-consistent solution of the Bloch-Horowitz (BH) equation\cite{Bloch:1958}. This
method was successfuly used to calculate the binding energy of $^3$He.
However, both of these works to date (\emph{i.e.} LS and BH
procedures) involve handling the excluded-space physics in a very
large but still finite Hilbert space.  This approximation introduces
an extra dependence to the three-body binding on the size of the
excluded space.  Such dependence can only be removed by a true
summation over all excluded high-energy modes (i.e. excluded space of
infinte size).  Though this dependence is small due to the fact that
results of these previous works are quite accurate (both groups examined the
convergence of their results as a function of the size of the excluded
space), the rate of convergence clearly depends in detail on how
singular the underlying NN interaction is.  Here we develop a method
for calculating the effective three-body interaction that introduces
no such high-momentum cutoff, but instead truly integrates over all
high-energy modes.  The accuracy of our results is only limited by our
numerical precision.  Thus it can be applied with equal confidence to
any underlying NN reaction, regardless of that potential's description
of short-range physics.  This interaction is used in the simplest test
cases, $^3$He/$^3$H, and the results are compared with those that would
result from evaluating the effective interaction only at the two-body
level. Our treatment is based on solving the BH
equation as well, which produces a state-dependent (and thus
excitation-energy-dependent) Hermitian interaction,
\begin{equation}\label{eqn:BH} 
H_{eff}(E)=P\{H+H\frac{1}{E-QH}QH\}P,
\end{equation}
where
\begin{eqnarray*}
H &=& \sum_{i<j}^A (T_{ij}+V_{ij}) \\ &=& \sum_i^A T_i+\sum_{i<j}^A V_{ij}-\frac{P_{CM}^2}{2M_{A}}.
\end{eqnarray*}
Here the intrinsic Hamiltonian $H$ is obtained by summing the relative
kinetic energy and potential energy operators $T_{ij}$ and $V_{ij}$
over all nucleon pairs $i$ and $j$.  The relative kinetic energy is
found by subtracting the center-of-mass (CM) kinetic energy from the
total kinetic energy obtained by summing $T_i$ over all nucleons.
$M_A$ is the total mass of the A-body system, $Q$ is the excluded
space projection operator, and $P=1-Q$ is the included-space
projection operator that defines the finite, low-energy space in which
a direct diagonalization will be done, once $H_{eff}$ is obtained. As
$E$ is the desired eigenvalue, which is not known $a$ $priori$, the
equations must be solved self-consistently.  In the calculations
reported here, the Av18 potential is used as the bare $NN$ interaction
and no bare three-body interaction is included.

Because our application is to $^3$He/$^3$H, rather than to a heavier
nucleus, we will have solved the effective interaction at the A-body
level.  Thus we are not testing the assumption of a cluster expansion
here.  Instead, our focus is on the technique for solving the
effective interaction, and its relation to the approximate two-body
result. Using techniques familiar from Faddeev calculations, we first
determine an effective two-body-like interaction,
$V_{12,eff}^{(2+1)}$, from the scattering $t$-matrix. We find that the
induced effective three-body interaction can be evaluated
perturbatively by iterating on $V_{12,eff}^{(2+1)}$.  For a
certain range of oscillator parameters $b$, convergence is achieved
even for very small included spaces, including the limiting
$0\hbar\omega$ space.

We also develop a non-perturbative method. By rearranging the terms of
the original expansion for the three-body interaction, the series can
be summed exactly through numerical solution of an integral equation.
This method is very accurate and efficient for any included space and
any oscillator parameter.

In Section~\ref{preliminaries} we generalize the two-body formulation
presented in Ref.\cite{Haxton:2002kb}.  Section~\ref{sect:perturbative}
describes our perturbative treatment of the effective interaction,
while Section~\ref{sect:nonperturbative} presents the nonperturbative
solution via an integral equation.  The associated discussions focus
on qualitative issues, with derivations reserved for the appendices.
We discuss the results and the need for a simple test of the cluster
assumption (by applying current results to $^4$He) in
Section~\ref{sect:conclusion}.

\section{Preliminaries: Two-body Review\label{preliminaries}}

To guarantee that $H_{eff}$, like the bare $H$, is translationally
invariant one can work in a complete basis of three-nucleon HO Slater
determinants that includes all configurations with quanta
$N\hbar\omega < \Lambda\hbar\omega$.  Such a basis, in traditional
independent-particle shell-model coordinates, is overcomplete,
consisting of subspaces characterized by the eigenvalues of the
center-of-mass Hamiltonian $H_{CM}$ that do not interact via
$H_{eff}$.  For $A\leq$4, it is convenient to avoid this
overcompleteness by using a Jacobi basis, which reduces the $A$-body
intrinsic-Hamiltonian problem to an ($A$-1)-body problem. This is the
choice we make here.  The Jacobi included-space ($P$) basis
corresponding to the HO shell-model basis described above is thus
simple, consisting of all relative-coordinate configurations with
$N\hbar\omega < \Lambda\hbar\omega$. (We note that for $A>$4, the
single-particle basis is the more efficient basis for many-body
calculations.  However, an $H_{eff}$ developed in a Jacobi basis can
be easily transformed to single-particle coordinates for use in
shell-model-inspired diagonalizations.)

Such included spaces for the two-nucleon system are simple to
construct.  For example, a $\Lambda$=2 included space consists of
spatial wave functions of the form $|n=0,l=0>$, $|n=1,l=0>$, and
$|n=0,l=2>$, coupled to spin and isospin wave functions to maintain
antisymmetry.  This requires $l+s+t$ to be odd.  The corresponding
construction of fully anti-symmetrized $A$-body wave functions is less
trivial \cite{Moshinsky:1969}.  Here we use the codes described in 
Ref.\cite{Navratil:1999pw} to generate these for $A$=3.  In
Table~\ref{tab:dimension} dimensions of various $A$=3 included spaces
are given as a function of $\Lambda$.

For the three-body system one can obtain accurate results by
diagonalizing the bare $H$ in a Jacobi basis, provided $\Lambda$ is
made sufficiently large.  This was done in Ref.\cite{Haxton:1999vg} for
the Av18 potential with $\Lambda$=60, resulting in a binding energy
accurate to $\sim$ 20 keV.  The motivation for solving this problem
for small $\Lambda$ using effective interactions is to explore
techniques that might be more feasible for larger $A$, where the
model-space growth analogous to Table~\ref{tab:dimension} will be much
steeper.

One goal of the current work is to find techniques that might make the
integration over the excluded space more tractable.  In our earlier
work on the two-body system\cite{Haxton:2002kb} we found that this
integration is difficult at both long and short distances.  We found
it convenient in Ref.\cite{Haxton:2002kb} to remove the long-distance
difficulties, the pathologies associated with the overbinding of the
HO, at the outset. As the overbinding of the HO becomes arbitrarily
large as $r$ increases, part of the tail of the wave function remains
unresolved in any finite-basis treatment-- though numerically the
contribution of the tail becomes increasingly unimportant in direct
diagonalizations as more shells are added.  (This contrasts with the
short-range problem, as Av18 and other modern potentials are regulated
at short distance by some assumed functional form. Such potentials can
be fully resolved in finite bases, provided $\Lambda$ is larger than
the scale implicit in that functional form.)  In Ref.\cite{Haxton:2002kb}
the long-range behavior was handled by the following rearrangement of
Eq.~(\ref{eqn:BH})
\begin{equation}\label{eqn:BHre}
H_{eff}=P\left\{\frac{E}{E-TQ}[T_{eff}+V_{eff}]\frac{E}{E-QT}\right\}P,
\end{equation}
where
\begin{eqnarray}
T_{eff} &=& T+T\frac{-1}{E}QT, \label{eqn:Teff}\\ V_{eff} &=&
V+V\frac{1}{E-QH}QV.\label{eqn:Veff}
\end{eqnarray}
Here $T$ and $V$ are shorthand for $\sum_{i<j}^A T_{ij}$ and
$\sum_{i<j}^A V_{ij}$, respectively.  Due to space restrictions
Eq.~\ref{eqn:BHre} was stated in Ref.\cite{Haxton:2002kb} without
derivation.  Hence we show its derivation for completeness in
Appendix~\ref{app:bhfinal}\footnote{Note that $T_{eff}$ appearing in
  Ref.\cite{Haxton:2002kb} differs from the one shown in
  Eq.~\ref{eqn:Teff}.  However, both expressions, from an operator
  standpoint, are completely equivalent.}.

To calculate included-space matrix elements of Eq.~\ref{eqn:BHre}, it
was first convenient to define the states
\begin{equation}\label{eqn:tilde_state}
\widetilde{|\Omega>}=\frac{E}{E-QT}P|\Omega>=\frac{E}{E-QT}|\Omega>,
\end{equation}
where $|\Omega>$ is some state that resides in the included space
(\emph{i.e.} $|\Omega>\in P$).  With this definition, matrix elements
of Eq.~\ref{eqn:BHre} became
\begin{equation}\label{eqn:Heff_ME}
<\Omega_f|H_{eff}|\Omega_i>=\widetilde{<\Omega_f|}T_{eff}+V_{eff}\widetilde{|\Omega_i>}.
\end{equation}
Our expansion of $H_{eff}$ was formed by expanding the resolvent of
$V_{eff}$ (see Eq.~\ref{eqn:Veff}) in powers of $QV$:
\begin{equation}\label{eqn:resolvent_expansion}
\frac{1}{E-QH}Q=\frac{1}{E-QT}Q+\frac{1}{E-QT}QV\frac{1}{E-QT}Q+\ldots
\end{equation}
An iterative procedure was then used to calculate included-space
matrix elements of this expansion.  At each order we solved for the
self-consistent energy $E$. 

\begin{figure*}
\begingroup%
  \makeatletter%
  \newcommand{\GNUPLOTspecial}{%
    \@sanitize\catcode`\%=14\relax\special}%
  \setlength{\unitlength}{0.1bp}%
\begin{picture}(3600,3023)(0,0)%
\includegraphics{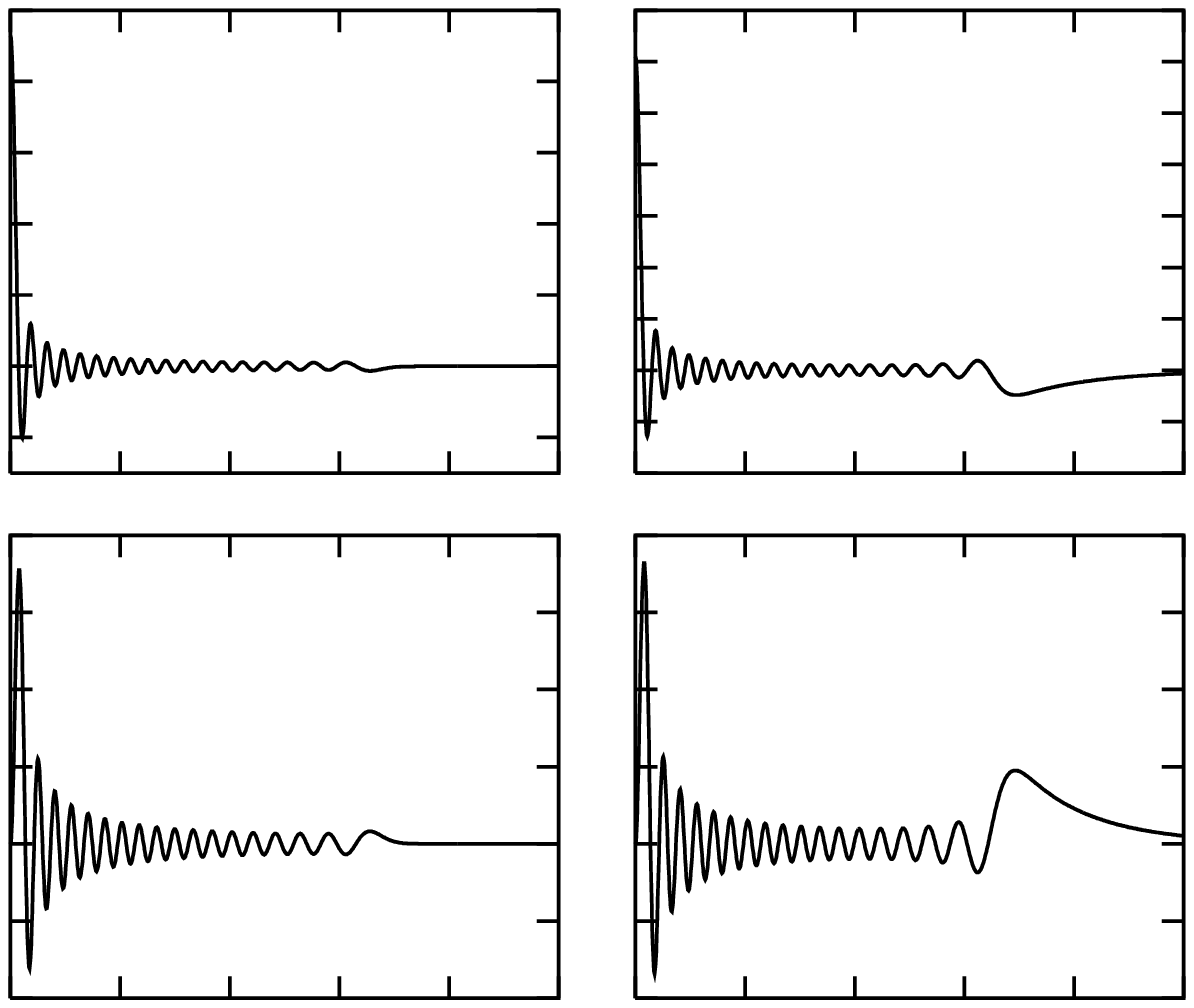}
\fontsize{9}{\baselineskip}\selectfont
\put(3217,246){\makebox(0,0)[l]{$r$ (fm)}}%
\put(2712,1021){\makebox(0,0)[l]{ $b=1.0$ fm}}%
\put(2712,1111){\makebox(0,0)[l]{ $\Lambda\hbar\omega=70\hbar\omega$}}%
\put(2712,1246){\makebox(0,0)[l]{$<r|\frac{E}{E-QT}|34\ D>$}}%
\put(2743,23){\makebox(0,0){(d)}}%
\put(3533,90){\makebox(0,0){ 25}}%
\put(3217,90){\makebox(0,0){ 20}}%
\put(2901,90){\makebox(0,0){ 15}}%
\put(2586,90){\makebox(0,0){ 10}}%
\put(2270,90){\makebox(0,0){ 5}}%
\put(1954,90){\makebox(0,0){ 0}}%
\put(1932,1468){\makebox(0,0)[r]{ 0.4}}%
\put(1932,1246){\makebox(0,0)[r]{ 0.3}}%
\put(1932,1024){\makebox(0,0)[r]{ 0.2}}%
\put(1932,801){\makebox(0,0)[r]{ 0.1}}%
\put(1932,579){\makebox(0,0)[r]{ 0}}%
\put(1932,357){\makebox(0,0)[r]{-0.1}}%
\put(1932,135){\makebox(0,0)[r]{-0.2}}%
\put(1417,246){\makebox(0,0)[l]{$r$ (fm)}}%
\put(912,1021){\makebox(0,0)[l]{ $b=1.0$ fm}}%
\put(912,1111){\makebox(0,0)[l]{ $\Lambda\hbar\omega=70\hbar\omega$}}%
\put(912,1246){\makebox(0,0)[l]{$<r|34\ D>$}}%
\put(943,23){\makebox(0,0){(c)}}%
\put(1733,90){\makebox(0,0){ 25}}%
\put(1417,90){\makebox(0,0){ 20}}%
\put(1101,90){\makebox(0,0){ 15}}%
\put(786,90){\makebox(0,0){ 10}}%
\put(470,90){\makebox(0,0){ 5}}%
\put(154,90){\makebox(0,0){ 0}}%
\put(132,1468){\makebox(0,0)[r]{ 0.8}}%
\put(132,1246){\makebox(0,0)[r]{ 0.6}}%
\put(132,1024){\makebox(0,0)[r]{ 0.4}}%
\put(132,801){\makebox(0,0)[r]{ 0.2}}%
\put(132,579){\makebox(0,0)[r]{ 0}}%
\put(132,357){\makebox(0,0)[r]{-0.2}}%
\put(132,135){\makebox(0,0)[r]{-0.4}}%
\put(3217,1758){\makebox(0,0)[l]{$r$ (fm)}}%
\put(2712,2533){\makebox(0,0)[l]{ $b=1.0$ fm}}%
\put(2712,2623){\makebox(0,0)[l]{ $\Lambda\hbar\omega=70\hbar\omega$}}%
\put(2712,2758){\makebox(0,0)[l]{$<r|\frac{E}{E-QT}|35\ S>$}}%
\put(2743,1535){\makebox(0,0){(b)}}%
\put(3533,1602){\makebox(0,0){ 25}}%
\put(3217,1602){\makebox(0,0){ 20}}%
\put(2901,1602){\makebox(0,0){ 15}}%
\put(2586,1602){\makebox(0,0){ 10}}%
\put(2270,1602){\makebox(0,0){ 5}}%
\put(1954,1602){\makebox(0,0){ 0}}%
\put(1932,2980){\makebox(0,0)[r]{ 1.4}}%
\put(1932,2832){\makebox(0,0)[r]{ 1.2}}%
\put(1932,2684){\makebox(0,0)[r]{ 1}}%
\put(1932,2536){\makebox(0,0)[r]{ 0.8}}%
\put(1932,2388){\makebox(0,0)[r]{ 0.6}}%
\put(1932,2239){\makebox(0,0)[r]{ 0.4}}%
\put(1932,2091){\makebox(0,0)[r]{ 0.2}}%
\put(1932,1943){\makebox(0,0)[r]{ 0}}%
\put(1932,1795){\makebox(0,0)[r]{-0.2}}%
\put(1932,1647){\makebox(0,0)[r]{-0.4}}%
\put(1417,1750){\makebox(0,0)[l]{$r$ (fm)}}%
\put(912,2550){\makebox(0,0)[l]{ $b=1.0$ fm}}%
\put(912,2640){\makebox(0,0)[l]{ $\Lambda\hbar\omega=70\hbar\omega$}}%
\put(912,2775){\makebox(0,0)[l]{$<r|35\ S>$}}%
\put(943,1535){\makebox(0,0){(a)}}%
\put(1733,1602){\makebox(0,0){ 25}}%
\put(1417,1602){\makebox(0,0){ 20}}%
\put(1101,1602){\makebox(0,0){ 15}}%
\put(786,1602){\makebox(0,0){ 10}}%
\put(470,1602){\makebox(0,0){ 5}}%
\put(154,1602){\makebox(0,0){ 0}}%
\put(132,2980){\makebox(0,0)[r]{ 2.5}}%
\put(132,2775){\makebox(0,0)[r]{ 2}}%
\put(132,2570){\makebox(0,0)[r]{ 1.5}}%
\put(132,2365){\makebox(0,0)[r]{ 1}}%
\put(132,2160){\makebox(0,0)[r]{ 0.5}}%
\put(132,1955){\makebox(0,0)[r]{ 0}}%
\put(132,1750){\makebox(0,0)[r]{-0.5}}%
\end{picture}%
\endgroup
\caption{Effect of $\frac{E}{E-QT}$ operator on a two-body endshell HO state for
  $70\hbar \omega$ included space.  Plot a (c) shows the un-corrected HO
  state, while plot b (d) shows the ensuing endshell corrections.
  Note the large corrections at large-$r$.\label{fig:endshell}}
\end{figure*}

The rearrangement of Eq.~(\ref{eqn:BHre}) was applied to the deuteron
in Ref.\cite{Haxton:2002kb} and led to excellent results in
perturbation theory for suitably chosen oscillator parameters $b$
(only few orders of $QV$ were needed to reproduce the deuteron binding
energy to within one keV). Figure~\ref{fig:endshell} shows that
$\frac{E}{E-QT}$ acting on the HO state $|\Omega>$ drastically changes
that state's large-$r$ behavior: The asymptotic behavior of
$\widetilde{|\Omega>}$ is $e^{-\gamma r}/\gamma r$, where
$\gamma=\sqrt{mE}$, $m$ being the mass of the particle\cite{Luu:2003}.
As this is the proper fall-off, the overconfining effects of the HO
are thus repaired.  The effects are important for the deuteron, which
has an extended wave function because of its small binding energy.
The operator $\frac{E}{E-QT}$ modifies only those HO states that
reside in the last shell of the included space.  For all other states
within the included space, the operator is identical to unity.  The
``endshell-corrected'' states thus control all of the proper
asymptotic behavior.

The resummation of the kinetic energy operator allows one to adjust
basis wave functions to absorb much of the short-range behavior of the
$NN$ interaction into the included space, leaving a weak residual
interaction that can be handled perturbatively.  By adjusting the
oscillator parameter $b$ to low values ($\sim 0.4-0.5$ fm), the
residual Q-space contributions to the binding energy due to
$V\frac{1}{E-QH}QV$ can be evaluated to within $\sim$ one keV in
third-order perturbation theory, even for very small included spaces.
This is only possible because the kinetic energy has been summed to
all orders independent of $b$, thus guaranteeing that the choice of a
small $b$ will not alter the wave function at large $r$.

The optimal $b$ presumably provides the best resolution of the hard
core without substantially altering the ability of the basis to
reconstruct the intermediate-range potential.

In the next section we will explore a similar expansion for
$^3$H/$^3$He. We find modifications are necessary to account for the
disparate length scales that come about from the inclusion of a third
particle.  Before starting this discussion, we end this section by
presenting compact expressions for the operators $\frac{E}{E-QT}P$ and
$\frac{1}{E-QT}Q$.  Such operators appear frequently in subsequent
formulae.

\subsection{$\frac{E}{E-QT}P$ and $\frac{1}{E-QT}Q$ operators} 

The operator $\frac{1}{E-QT}$ closely resembles the free particle
propagator. However, due to the projection operator Q in the
denominator, calculating matrix elements of $\frac{E}{E-QT}P$ requires
a little ingenuity.  Here we show that the excluded-space
Green's function can be expanded in terms of the much simpler
full Green's function and operators within the included space
that can be inverted easily.  Following Ref.\cite{Krenciglowa:1976}, we first
consider
\begin{multline}
PT\frac{E}{E-QT}P=PT\{\frac{E}{E-T}P
  -\frac{1}{E-T}PT\frac{E}{E-QT}\}P,
\label{eqn:step1}
\end{multline}
where $Q=1-P$. Collecting terms in this ``Schwinger-Dyson'' form 
and noting that $P^2=P$ gives
\begin{equation}
\begin{aligned}
P&\left\{1+T\frac{1}{E-T}\right\}P\ PT\frac{E}{E-QT}P=PT\frac{E}{E-T}P,\\
\Rightarrow& PT\frac{E}{E-QT}P=\left\{P\frac{E}{E-T}P\right\}^{-1} \
PT\frac{E}{E-T}P.
\end{aligned}\label{eqn:step2}
\end{equation}
The last term in Eq.~\ref{eqn:step2}, $PT\frac{E}{E-T}P$, can be
rewritten
\begin{displaymath}
PT\frac{E}{E-T}P=EP\left(P\frac{E}{E-T}P-1\right)P.
\end{displaymath}
Plugging the above equation into Eq.~\ref{eqn:step2} finally gives
\begin{equation}
PT\frac{E}{E-QT}P=P\left(E-\left\{P\frac{1}{E-T}P\right\}^{-1}\right)P.
\label{eqn:step3}
\end{equation}
Similarly, the ``Schwinger-Dyson'' form of $\frac{E}{E-QT}P$ is
\begin{equation}
\frac{E}{E-QT}P=\frac{E}{E-T}P-\frac{1}{E-T}\left[PT\frac{E}{E-QT}P\right].
\end{equation}
Using Eq.~(\ref{eqn:step3}) then yields
\begin{equation}
\frac{E}{E-QT}P=\frac{1}{E-T}\left\{P\frac{1}{E-T}P\right\}^{-1}P.
\label{eqn:finalstep}
\end{equation}
This expression is relatively simple to use.  The free particle
propagator, $\frac{1}{E-T}$, is known analytically.  Indeed, its form
is diagonal in momentum space.  The operator $P\frac{1}{E-T}P$
represents a matrix composed of included-space overlaps of the free
particle propagator.  Since we work within small included spaces,
inverting this matrix is not difficult.  The included-space matrix
elements are easy to calculate, as analytic expressions for
$<\Omega_f|\frac{1}{E-T}|\Omega_i>$ for any $A$-body system exists.
These expressions involve multiple sums over hypergeometric functions
and gamma functions.  In Appendix~\ref{sect:3bodyETME} we give the
relevant expression for the three-body system.

To derive an analogous expression for $\frac{1}{E-QT}Q$, we first
consider
\begin{equation}
\begin{aligned}
\frac{1}{E-QT}=& \frac{1}{E-T}-\frac{1}{E-QT}PT\frac{1}{E-T}\\
 = \frac{1}{E-T}-&\frac{1}{E-T}\left\{P\frac{1}{E-T}P\right\}^{-1}
  \frac{T}{E}\frac{1}{E-T},
\end{aligned}\label{eqn:EminusQT}
\end{equation}
where we have substituted Eq.~\ref{eqn:finalstep} in the second line
above.  With the use of Eqs.~\ref{eqn:finalstep}
and~\ref{eqn:EminusQT}, $\frac{1}{E-QT}Q$ becomes
\begin{equation}
\begin{aligned}
\frac{1}{E-QT}Q=&\frac{1}{E-QT}(1-P)\\
=\frac{1}{E-T}-&\frac{1}{E-T}\left\{P\frac{1}{E-T}P\right\}^{-1}
\frac{1}{E-T}.
\end{aligned}\label{eqn:EminusQTQ}
\end{equation}
The physical interpretation of Eq.~\ref{eqn:EminusQTQ} is simple: the
term on the \emph{LHS} represents free propagation in the excluded
($Q$) space, while the first term on the \emph{RHS} (bottom line)
represents free propagation in the full ($P+Q$) space and the second
term subtracts off the contribution coming from included-space ($P$)
propagation.

It is convenient to make the following definitions:
\begin{eqnarray}
G_0&=&\frac{1}{E-T},\label{eqn:G0}\\
\Gamma_0^{-1}&=&\left\{P\frac{1}{E-T}P\right\}^{-1}.\label{eqn:invGamma0}
\end{eqnarray}
With these definitions, Eqs.~\ref{eqn:finalstep}
and~\ref{eqn:EminusQTQ} become
\begin{eqnarray*}
\frac{E}{E-QT}P&=&G_0\Gamma_0^{-1},\\
\frac{1}{E-QT}Q&=&G_0-G_0\Gamma_0^{-1}G_0.
\end{eqnarray*}

Finally, with the expressions above, we can express our perturbative
expansion in a concise form,
\begin{equation}\label{eqn:bhfinal}
\begin{aligned}
H_{eff}=&E-\Gamma_0^{-1}+\Gamma_0^{-1}\Gamma_1\Gamma_0^{-1}\\
+&\Gamma_0^{-1}\left\{\Gamma_2-\Gamma_1\Gamma_0^{-1}\Gamma_1\right\}\Gamma_0^{-1}
+&\ldots,
\end{aligned}
\end{equation}
where
\begin{equation}\label{eqn:Gamma_n}
\begin{aligned}
\Gamma_1=&PG_0VG_0P\\
\Gamma_2=&PG_0VG_0VG_0P\\
\vdots\quad&\quad\quad\vdots\\
\Gamma_n=&PG_0\left(VG_0\right)^nP.
\end{aligned}
\end{equation}
In the expansion given by Eq.~\ref{eqn:bhfinal}, $E-\Gamma_0^{-1}$
corresponds to $\frac{E}{E-TQ}T_{eff}\frac{E}{E-QT}$, while
$\Gamma_0^{-1}\Gamma_1\Gamma_0^{-1}$ corresponds to
$\frac{E}{E-TQ}V\frac{E}{E-QT}$, and so on.  As we stressed in
Ref.\cite{Haxton:2002kb}, matrix elements of the operators shown in
Eq.~\ref{eqn:Gamma_n} can be simply calculated using a recursive
procedure.

\section{A Perturbative Three-body calculation\label{sect:perturbative}}
\begin{turnpage}
\begin{figure*}
\begingroup%
  \makeatletter%
  \newcommand{\GNUPLOTspecial}{%
    \@sanitize\catcode`\%=14\relax\special}%
  \setlength{\unitlength}{0.1bp}%
\begin{picture}(3600,4320)(0,0)%
\includegraphics{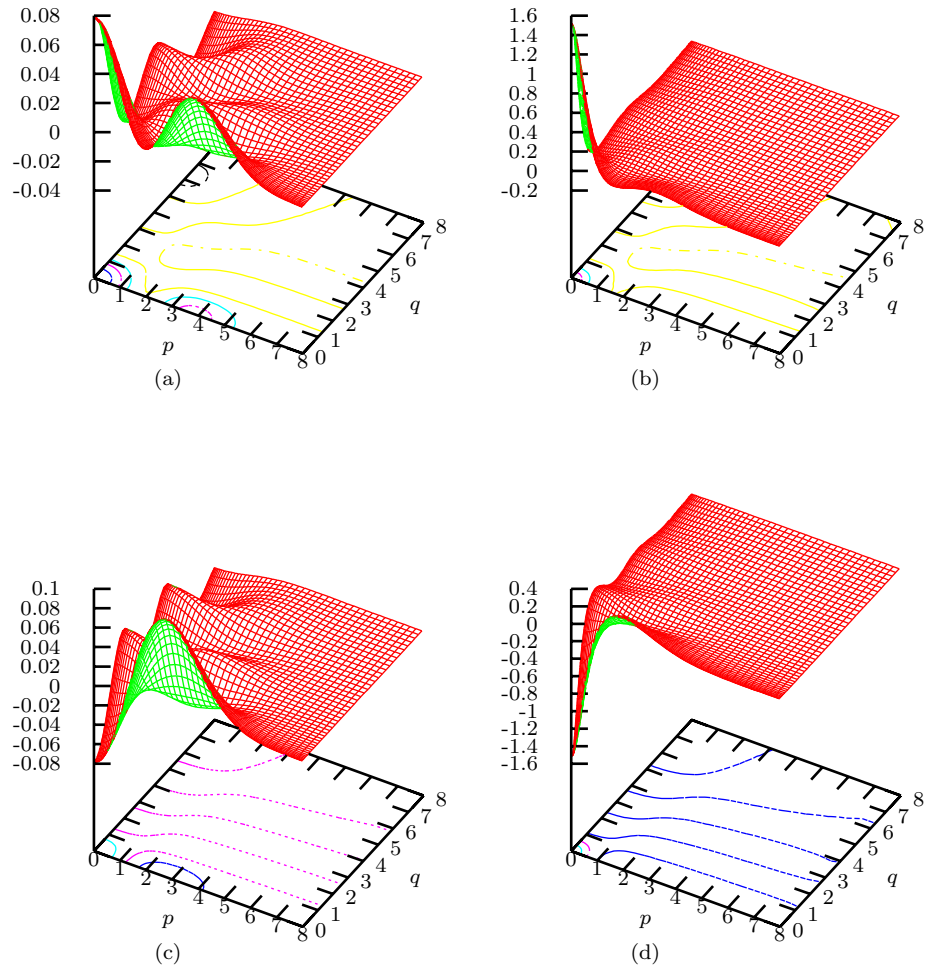}
\fontsize{8}{\baselineskip}\selectfont
\put(1952,1489){\makebox(0,0)[r]{ 0.4}}%
\put(1952,1423){\makebox(0,0)[r]{ 0.2}}%
\put(1952,1357){\makebox(0,0)[r]{ 0}}%
\put(1952,1292){\makebox(0,0)[r]{-0.2}}%
\put(1952,1226){\makebox(0,0)[r]{-0.4}}%
\put(1952,1160){\makebox(0,0)[r]{-0.6}}%
\put(1952,1094){\makebox(0,0)[r]{-0.8}}%
\put(1952,1028){\makebox(0,0)[r]{-1}}%
\put(1952,962){\makebox(0,0)[r]{-1.2}}%
\put(1952,896){\makebox(0,0)[r]{-1.4}}%
\put(1952,830){\makebox(0,0)[r]{-1.6}}%
\put(3291,391){\makebox(0,0){$q$}}%
\put(3339,696){\makebox(0,0)[l]{ 8}}%
\put(3282,634){\makebox(0,0)[l]{ 7}}%
\put(3225,572){\makebox(0,0)[l]{ 6}}%
\put(3168,511){\makebox(0,0)[l]{ 5}}%
\put(3112,449){\makebox(0,0)[l]{ 4}}%
\put(3055,387){\makebox(0,0)[l]{ 3}}%
\put(2998,325){\makebox(0,0)[l]{ 2}}%
\put(2941,264){\makebox(0,0)[l]{ 1}}%
\put(2884,202){\makebox(0,0)[l]{ 0}}%
\put(2358,115){\makebox(0,0){(d)}}%
\put(2358,235){\makebox(0,0){$p$}}%
\put(2852,186){\makebox(0,0){ 8}}%
\put(2754,221){\makebox(0,0){ 7}}%
\put(2655,257){\makebox(0,0){ 6}}%
\put(2557,293){\makebox(0,0){ 5}}%
\put(2458,328){\makebox(0,0){ 4}}%
\put(2359,364){\makebox(0,0){ 3}}%
\put(2261,400){\makebox(0,0){ 2}}%
\put(2162,435){\makebox(0,0){ 1}}%
\put(2064,471){\makebox(0,0){ 0}}%
\put(152,1489){\makebox(0,0)[r]{ 0.1}}%
\put(152,1416){\makebox(0,0)[r]{ 0.08}}%
\put(152,1343){\makebox(0,0)[r]{ 0.06}}%
\put(152,1270){\makebox(0,0)[r]{ 0.04}}%
\put(152,1196){\makebox(0,0)[r]{ 0.02}}%
\put(152,1123){\makebox(0,0)[r]{ 0}}%
\put(152,1050){\makebox(0,0)[r]{-0.02}}%
\put(152,977){\makebox(0,0)[r]{-0.04}}%
\put(152,904){\makebox(0,0)[r]{-0.06}}%
\put(152,830){\makebox(0,0)[r]{-0.08}}%
\put(1491,391){\makebox(0,0){$q$}}%
\put(1539,696){\makebox(0,0)[l]{ 8}}%
\put(1482,634){\makebox(0,0)[l]{ 7}}%
\put(1425,572){\makebox(0,0)[l]{ 6}}%
\put(1368,511){\makebox(0,0)[l]{ 5}}%
\put(1312,449){\makebox(0,0)[l]{ 4}}%
\put(1255,387){\makebox(0,0)[l]{ 3}}%
\put(1198,325){\makebox(0,0)[l]{ 2}}%
\put(1141,264){\makebox(0,0)[l]{ 1}}%
\put(1084,202){\makebox(0,0)[l]{ 0}}%
\put(558,115){\makebox(0,0){(c)}}%
\put(558,235){\makebox(0,0){$p$}}%
\put(1052,186){\makebox(0,0){ 8}}%
\put(954,221){\makebox(0,0){ 7}}%
\put(855,257){\makebox(0,0){ 6}}%
\put(757,293){\makebox(0,0){ 5}}%
\put(658,328){\makebox(0,0){ 4}}%
\put(559,364){\makebox(0,0){ 3}}%
\put(461,400){\makebox(0,0){ 2}}%
\put(362,435){\makebox(0,0){ 1}}%
\put(264,471){\makebox(0,0){ 0}}%
\put(1952,3649){\makebox(0,0)[r]{ 1.6}}%
\put(1952,3576){\makebox(0,0)[r]{ 1.4}}%
\put(1952,3503){\makebox(0,0)[r]{ 1.2}}%
\put(1952,3430){\makebox(0,0)[r]{ 1}}%
\put(1952,3356){\makebox(0,0)[r]{ 0.8}}%
\put(1952,3283){\makebox(0,0)[r]{ 0.6}}%
\put(1952,3210){\makebox(0,0)[r]{ 0.4}}%
\put(1952,3137){\makebox(0,0)[r]{ 0.2}}%
\put(1952,3064){\makebox(0,0)[r]{ 0}}%
\put(1952,2990){\makebox(0,0)[r]{-0.2}}%
\put(3291,2551){\makebox(0,0){$q$}}%
\put(3339,2856){\makebox(0,0)[l]{ 8}}%
\put(3282,2794){\makebox(0,0)[l]{ 7}}%
\put(3225,2732){\makebox(0,0)[l]{ 6}}%
\put(3168,2671){\makebox(0,0)[l]{ 5}}%
\put(3112,2609){\makebox(0,0)[l]{ 4}}%
\put(3055,2547){\makebox(0,0)[l]{ 3}}%
\put(2998,2485){\makebox(0,0)[l]{ 2}}%
\put(2941,2424){\makebox(0,0)[l]{ 1}}%
\put(2884,2362){\makebox(0,0)[l]{ 0}}%
\put(2358,2275){\makebox(0,0){(b)}}%
\put(2358,2395){\makebox(0,0){$p$}}%
\put(2852,2346){\makebox(0,0){ 8}}%
\put(2754,2381){\makebox(0,0){ 7}}%
\put(2655,2417){\makebox(0,0){ 6}}%
\put(2557,2453){\makebox(0,0){ 5}}%
\put(2458,2488){\makebox(0,0){ 4}}%
\put(2359,2524){\makebox(0,0){ 3}}%
\put(2261,2560){\makebox(0,0){ 2}}%
\put(2162,2595){\makebox(0,0){ 1}}%
\put(2064,2631){\makebox(0,0){ 0}}%
\put(152,3649){\makebox(0,0)[r]{ 0.08}}%
\put(152,3539){\makebox(0,0)[r]{ 0.06}}%
\put(152,3430){\makebox(0,0)[r]{ 0.04}}%
\put(152,3320){\makebox(0,0)[r]{ 0.02}}%
\put(152,3210){\makebox(0,0)[r]{ 0}}%
\put(152,3100){\makebox(0,0)[r]{-0.02}}%
\put(152,2990){\makebox(0,0)[r]{-0.04}}%
\put(1491,2551){\makebox(0,0){$q$}}%
\put(1539,2856){\makebox(0,0)[l]{ 8}}%
\put(1482,2794){\makebox(0,0)[l]{ 7}}%
\put(1425,2732){\makebox(0,0)[l]{ 6}}%
\put(1368,2671){\makebox(0,0)[l]{ 5}}%
\put(1312,2609){\makebox(0,0)[l]{ 4}}%
\put(1255,2547){\makebox(0,0)[l]{ 3}}%
\put(1198,2485){\makebox(0,0)[l]{ 2}}%
\put(1141,2424){\makebox(0,0)[l]{ 1}}%
\put(1084,2362){\makebox(0,0)[l]{ 0}}%
\put(558,2275){\makebox(0,0){(a)}}%
\put(558,2395){\makebox(0,0){$p$}}%
\put(1052,2346){\makebox(0,0){ 8}}%
\put(954,2381){\makebox(0,0){ 7}}%
\put(855,2417){\makebox(0,0){ 6}}%
\put(757,2453){\makebox(0,0){ 5}}%
\put(658,2488){\makebox(0,0){ 4}}%
\put(559,2524){\makebox(0,0){ 3}}%
\put(461,2560){\makebox(0,0){ 2}}%
\put(362,2595){\makebox(0,0){ 1}}%
\put(264,2631){\makebox(0,0){ 0}}%
\end{picture}%
\endgroup
\caption{Effect of $\frac{E}{E-QT}$ operator on a particular three-body endshell HO state for
  $10\hbar \omega$ included space.  Plots a and c show the uncorrected
  wavefunctions for the $^1S_0$ and $^3S_1$ channels, respectively,
  while plots b and d show the ensuing corrections.  Note the large
  corrections at small-$p$, corresponding to large corrections at large-$r$.\label{fig:3bodyendshell}}
\end{figure*}
\end{turnpage}

Similarly, one might apply the expansion given by
Eqs.~(\ref{eqn:bhfinal}) and~(\ref{eqn:Gamma_n}) directly to
$^3$He/$^3$H.  As was done in the two-body system, we construct
endshell-corrected states to build in the correct aymptotic forms.
Figure~\ref{fig:3bodyendshell} shows momentum-space results for one
$10\hbar\omega$ endshell state.  As expected, the modified
wavefunction acquires a strong peak near zero-momentum, corresponding
to large-r corrections.

\begin{figure}
\begin{center}
\begin{picture}(0,0)%
\includegraphics{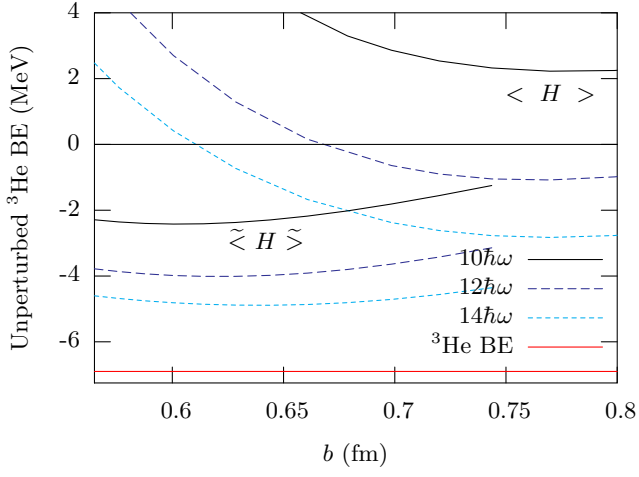}%
\end{picture}%
\setlength{\unitlength}{0.0200bp}%
\begin{picture}(12599,9180)(0,0)%
\put(1650,2426){\makebox(0,0)[r]{\strut{}-6}}%
\put(1650,3666){\makebox(0,0)[r]{\strut{}-4}}%
\put(1650,4907){\makebox(0,0)[r]{\strut{}-2}}%
\put(1650,6148){\makebox(0,0)[r]{\strut{} 0}}%
\put(1650,7389){\makebox(0,0)[r]{\strut{} 2}}%
\put(1650,8630){\makebox(0,0)[r]{\strut{} 4}}%
\put(3392,1100){\makebox(0,0){\strut{} 0.6}}%
\put(5488,1100){\makebox(0,0){\strut{} 0.65}}%
\put(7584,1100){\makebox(0,0){\strut{} 0.7}}%
\put(9679,1100){\makebox(0,0){\strut{} 0.75}}%
\put(11775,1100){\makebox(0,0){\strut{} 0.8}}%
\put(550,5140){\rotatebox{90}{\makebox(0,0){\strut{}Unperturbed $^3$He BE (MeV)}}}%
\put(6850,275){\makebox(0,0){\strut{}$b$ (fm)}}%
\put(9679,7079){\makebox(0,0)[l]{\strut{}$<\ H\ >$}}%
\put(4440,4287){\makebox(0,0)[l]{\strut{}$\widetilde{<}\ H\ \widetilde{>}$}}%
\put(9823,3977){\makebox(0,0)[r]{\strut{}$10\hbar\omega$}}%
\put(9823,3427){\makebox(0,0)[r]{\strut{}$12\hbar\omega$}}%
\put(9823,2877){\makebox(0,0)[r]{\strut{}$14\hbar\omega$}}%
\put(9823,2327){\makebox(0,0)[r]{\strut{}$^3$He BE}}%
\end{picture}%
\caption[Variational calculation on $^3$He for various included space sizes]
{Variational results for $^3$He for various included spaces.  The 
upper sets of lines correspond to calculations without endshell
corrections, while the lower sets have endshell corrections.\label{fig:he3variational}}
\end{center}
\end{figure}

To determine an optimal $b$ for subsequent calculations, $H$ is
minimized within the included space. The results are given in
Fig.~\ref{fig:he3variational} for several included spaces.  As
expected, the endshell states improve the binding of the three-body
system and lower the optimal $b$, relative to uncorrected HO results.
However, the improved binding energies are still far away from the
exact answers\footnote{Exact answers are taken from Faddeev
  calculations\cite{Nogga:2002qp}.}.  This contrasts with the
deuterium results (see Fig.2 of Ref.~\cite{Haxton:2002kb}), where
0th-order results using endshell states were accurate ( $\sim$ 50
keV), and became nearly exact in second- or third-order perturbation
theory.  The 0th order A=3 results underbind by $\sim 3$ MeV.

\begin{figure}
\begin{center}
\begin{picture}(0,0)%
\includegraphics{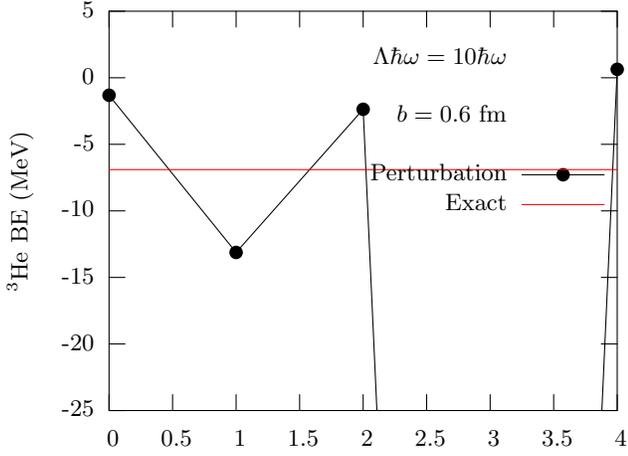}%
\end{picture}%
\setlength{\unitlength}{0.0200bp}%
\begin{picture}(12599,9180)(0,0)%
\put(1925,1100){\makebox(0,0)[r]{\strut{}-25}}%
\put(1925,2355){\makebox(0,0)[r]{\strut{}-20}}%
\put(1925,3610){\makebox(0,0)[r]{\strut{}-15}}%
\put(1925,4865){\makebox(0,0)[r]{\strut{}-10}}%
\put(1925,6120){\makebox(0,0)[r]{\strut{}-5}}%
\put(1925,7375){\makebox(0,0)[r]{\strut{} 0}}%
\put(1925,8630){\makebox(0,0)[r]{\strut{} 5}}%
\put(2200,550){\makebox(0,0){\strut{} 0}}%
\put(3397,550){\makebox(0,0){\strut{} 0.5}}%
\put(4594,550){\makebox(0,0){\strut{} 1}}%
\put(5791,550){\makebox(0,0){\strut{} 1.5}}%
\put(6988,550){\makebox(0,0){\strut{} 2}}%
\put(8184,550){\makebox(0,0){\strut{} 2.5}}%
\put(9381,550){\makebox(0,0){\strut{} 3}}%
\put(10578,550){\makebox(0,0){\strut{} 3.5}}%
\put(11775,550){\makebox(0,0){\strut{} 4}}%
\put(550,4865){\rotatebox{90}{\makebox(0,0){\strut{}$^3$He BE (MeV)}}}%
\put(9705,7752){\makebox(0,0)[r]{\strut{}$\Lambda\hbar\omega=10\hbar\omega$}}%
\put(9705,7202){\makebox(0,0)[r]{\strut{}}}%
\put(9705,6652){\makebox(0,0)[r]{\strut{}$b=0.6$ fm}}%
\put(9705,6102){\makebox(0,0)[r]{\strut{}}}%
\put(9705,5546){\makebox(0,0)[r]{\strut{}Perturbation}}%
\put(9705,4984){\makebox(0,0)[r]{\strut{}Exact}}%
\end{picture}%
\caption[$10\hbar\omega$ included space calculation for
$^3$He]{$10\hbar\omega$ included-space calculation for $^3$He 
using expansion given by Eq.~\ref{eqn:bhfinal}.\label{fig:10hwHe3Energy}}
\end{center}
\end{figure}

\begin{figure}                                                       
\begin{center}                                                       
\epsfig{file=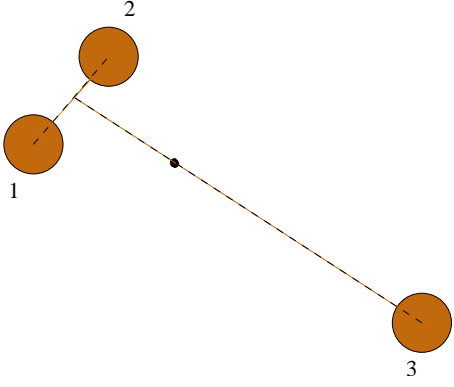,width=6cm}
\caption[Problematic configuration for three-body expansion]
{Problematic configuration for three-body expansion.  Such a
 configuration consists of two closely spaced particles (1 and 2), 
 while a third particle is separated far from the first
 two.\label{fig:efimov}}
\end{center}                                                        
\end{figure}

As Fig.~\ref{fig:10hwHe3Energy} shows, these results are not corrected
in perturbation theory.  Successive orders produce wildly oscillating
values for the binding energy, in contrast to the rapid convergence
found for the deuteron.  The difficulty is that a single distance
scale $b$ does not provide sufficient freedom to describe short-range
behavior governed by two relative cordinates.  Extended states such as
those of Fig.~\ref{fig:efimov} are problematic. Once one makes a
choice of Jacobi coordinates, the short-range behavior of some
two-body cluster will be difficult to describe.  For example, the
choice $(\vec{r}_3 - \vec{r}_2)/\sqrt{2}$ and $(2 \vec{r}_1 -
\vec{r}_3 - \vec{r}_2)/\sqrt{6}$ prevents one from adjusting $b$ to
account for the short-range interactions of nucleons 1 and 2.
Consequently, important hard-core interactions lie outside the
included space, leading to nonperturbative behavior.

\subsection{Invoking Faddeev Decompositions}

\begin{figure}
\begin{center}                                                       
\epsfig{file=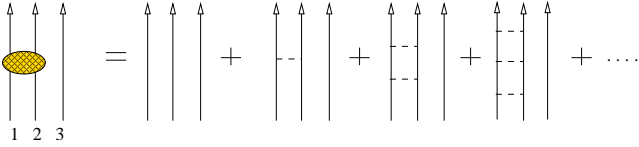,width=8.5cm}
\caption{Diagrammatic expansion of $\frac{1}{E-QT-QV_{12}}Q.$
  The dashed lines correspond to $V_{12}$ interactions.}
\label{fig:V12eff}                                                
\end{center}                                                 
\end{figure}

Because the deuteron was easily treated, it is possible to sum to all
orders the repeated excluded space scatterings of any two-nucleon
cluster.  Such a partial summation should account for the strong
repulsive interaction between any two coupled nucleons, leaving only
intermediate- and long-range interactions.  Such interactions can be
treated reasonably accurately within the included space using our
endshell-corrected states, leaving perturbative corrections.  This
partial summation corresponds to the Faddeev
decomposition\cite{Faddeev:1961su,Glockle:1983}.  The starting point
is again Eq.~\ref{eqn:Heff_ME},
\begin{equation*}
<\Omega_{f}|H_{eff}|\Omega{i}>=\widetilde{<\Omega_{f}|}T_{eff}+V_{eff}\widetilde{|\Omega_{i}>},
\end{equation*}
where we cast $V_{eff}$ in its integral form,
\begin{equation}\label{eqn:V3eff}
V_{eff}=V^3_{eff}=V+V\frac{1}{E-QT}QV^3_{eff}.
\end{equation}
Here the superscript $^3$ denotes that this is a three-body effective
interaction.  Now consider the state $V^3_{eff}\widetilde{|\ 
  \Omega_{\sigma}>}$ $\equiv$ $\widetilde{|\ \Psi_{\sigma}>}$, which
satisfies
\begin{equation}\label{eqn:psi_integralrel}
\widetilde{|\ \Psi_{\sigma}>}=V\widetilde{|\
\Omega_{\sigma}>}+V\frac{1}{E-QT}Q\widetilde{|\ \Psi_{\sigma}>}.
\end{equation}
This can be decomposed into Faddeev components
$\widetilde{|\Psi_{\sigma}>_{ij}}$, given by
\begin{equation}\label{eqn:psi_faddeev}
\begin{aligned}
\widetilde{|\ \Psi_{\sigma}>}_{12}=&V_{12}\widetilde{|\
\Omega_{\sigma}>}+V_{12}\frac{1}{E-QT}Q\widetilde{|\ \Psi_{\sigma}>},\\
\widetilde{|\ \Psi_{\sigma}>}_{23}=&V_{23}\widetilde{|\
\Omega_{\sigma}>}+V_{23}\frac{1}{E-QT}Q\widetilde{|\ \Psi_{\sigma}>},\\
\widetilde{|\ \Psi_{\sigma}>}_{31}=&V_{31}\widetilde{|\
\Omega_{\sigma}>}+V_{31}\frac{1}{E-QT}Q\widetilde{|\ \Psi_{\sigma}>},
\end{aligned}
\end{equation}
where
\begin{equation}
\widetilde{|\ \Psi_{\sigma}>}=\widetilde{|\
\Psi_{\sigma}>}_{12}+\widetilde{|\ \Psi_{\sigma}>}_{23}+\widetilde{|\
\Psi_{\sigma}>}_{31}.\label{eqn:psi_sum}
\end{equation}
Equations~\ref{eqn:psi_sum} and~\ref{eqn:psi_faddeev} show that the
Faddeev components require the solution of three coupled integral
equations.  For example, $\widetilde{|\ \Psi_{\sigma}>}_{12}$
satisfies
\begin{equation}\label{eqn:psi12_relation}
  \begin{aligned}
 \widetilde{|\ \Psi_{\sigma}>}_{12}&=V_{12}\widetilde{|\ \Omega_{\sigma}>}+\\
 V_{12}&\frac{1}{E-QT}Q\left(\widetilde{|\ \Psi_{\sigma}>}_{12}+\widetilde{|\ \Psi_{\sigma}>}_{23}+
   \widetilde{|\
 \Psi_{\sigma}>}_{31}\right).
\end{aligned}
\end{equation}
Equation~\ref{eqn:psi12_relation} can be inverted with respect to the
$\widetilde{|\ \Psi_{\sigma}>}_{12}$ Faddeev component, giving
\begin{equation}\label{eqn:psi12_relation2}
\begin{aligned}
 \widetilde{|\ \Psi_{\sigma}>}_{12}&=V^{(2+1)}_{12,eff}\widetilde{|\ \Omega_{\sigma}>}+\\
                 V^{(2+1)}_{12,eff}&\frac{1}{E-QT}Q\left(\widetilde{|\
                 \Psi_{\sigma}>}_{23}+\widetilde{|\ \Psi_{\sigma}>}_{31}\right),
\end{aligned}
\end{equation}
where
\begin{equation}\label{eqn:V12eff}
\begin{aligned}
V^{(2+1)}_{12,eff}=&\frac{1}{1-V_{12}\frac{1}{E-QT}Q}V_{12}\\
                  =&V_{12}+V_{12}\frac{1}{E-QT-QV_{12}}QV_{12}\\
                  =&V_{12}+V_{12}\frac{1}{E-QT}QV^{(2+1)}_{12,eff}.
\end{aligned}
\end{equation}
Analogous equations can be found for the remaining Faddeev components.
The superscript $^{(2+1)}$ indicates that the effective interaction
given by Eq.~\ref{eqn:V12eff} represents the sum of repeated potential
scatterings between the same two nucleons (in this case, nucleons 1
and 2), while the third nucleon remains a spectator, as illustrated in
Fig.~\ref{fig:V12eff}.

This expression is exactly the partial sum mentioned above.  Notice
that the middle expression in Eq.~\ref{eqn:V12eff} is very similar to
Eq.~\ref{eqn:Veff}.  Furthermore, the integral form of
$V^{(2+1)}_{12,eff}$ is similar to that of the
G-matrix\cite{Brueckner:1955}.  Yet there are subtleties that
differentiate the two.  We will return to this point later.  The
integral equation given by Eq.~\ref{eqn:V12eff} can also be solved
exactly in terms of the (2+1)-body $t$-matrices\cite{Glockle:1983}.
In Appendix~\ref{sect:exactV12eff} we show
\begin{equation}\label{eqn:exactV12eff}
V^{(2+1)}_{12,eff}=t_{12}-t_{12}G_0\left[\Gamma_0+\Gamma^{(2+1)}_{\infty}\right]^{-1}G_0t_{12},
\end{equation}
where
\begin{eqnarray}
t_{12}&=&V_{12}+V_{12}G_0t_{12} \label{eqn:t12def}\\
\Gamma^{(2+1)}_{\infty}&=&PG_0t_{12}G_0P\label{eqn:gamma21def},
\end{eqnarray}
and $\Gamma_0$ is given by Eq.~\ref{eqn:invGamma0}.  The physical
interpretation of Eq.~\ref{eqn:exactV12eff} is similar to the one
given below Eq.~\ref{eqn:EminusQT}: the first term on the \emph{RHS}
represents the effective interaction coming from repeated scatterings
in the full ($P+Q$) space, while the second term subtracts off the
contribution from the included ($P$) space.

The advantage in using the Faddeev decompositions comes from
exploiting particle exchange symmetries.  As the states $|\ 
\Omega_{\sigma}>$ are fully anti-symmetric, the individual Faddeev
components can be related to each other by simple permutation
operators\cite{Glockle:1983}.  In particular,
\begin{equation}\label{eqn:P_psi}
\begin{aligned}
\widetilde{|\ \Psi_{\sigma}>}_{23}=&P_{12}P_{23}\widetilde{|\
\Psi_{\sigma}>}_{12},\\ \widetilde{|\
\Psi_{\sigma}>}_{31}=&P_{12}P_{13}\widetilde{|\ \Psi_{\sigma}>}_{12},\\
\Rightarrow \widetilde{|\ \Psi_{\sigma}>}=&(1+\Pi)\widetilde{|\
\Psi_{\sigma}>}_{12},
\end{aligned}
\end{equation}
where the permutation operator $\Pi$ is defined as
$\Pi=P_{12}P_{13}+P_{12}P_{23}$.  The last expression in
Eq.~\ref{eqn:psi12_relation} then becomes
\begin{equation*}
\widetilde{|\ \Psi_{\sigma}>}_{12}=V^{(2+1)}_{12,eff}\widetilde{|\
\Omega_{\sigma}>}+ V^{(2+1)}_{12,eff}\frac{1}{E-QT}Q\Pi\widetilde{|\
\Psi_{\sigma}>}_{12},
\end{equation*}
which can be formally inverted with respect to $\widetilde{|\
  \Psi_{\sigma}>}_{12}$ to give
\begin{equation}\label{eqn:psi_summed}
 \widetilde{|\ \Psi_{\sigma}>}_{12}
 =\left(V^{(2+1)}_{12,eff}+V^{(3+0)}_{12,eff}\right)\widetilde{|\
 \Omega_{\sigma}>}.
\end{equation}
where
\begin{equation}\label{eqn:V30eff}
V^{(3+0)}_{12,eff}\equiv V^{(2+1)}_{12,eff}\frac{1}{E-QT-Q\Pi
V^{(2+1)}_{12,eff}}Q\Pi V^{(2+1)}_{12,eff}.
\end{equation}

\begin{figure}
\begin{center}                                                       
\epsfig{file=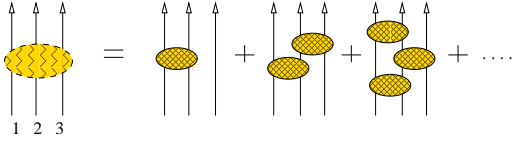,width=6.8cm}
\caption[Diagrammatic expansion of $\frac{1}{E-QH}Q$ for the three-body problem]
{Diagrammatic expansion of $\frac{1}{E-QH}Q$ for the three-body
  problem. The shaded ovals on the \emph{RHS} are given by Fig.~\ref{fig:V12eff}.  
 Aside from the first term on the \emph{RHS}, the remaining terms
 contribute to $V^{(3+0)}_{12,eff}$.}  
\label{fig:resolvent30}                                                
\end{center}                                                         
\end{figure}

Equation~\ref{eqn:V30eff} defines the induced three-body interaction
coming from repeated scatterings of $V^{(2+1)}_{12,eff}$ (see
Fig.~\ref{fig:resolvent30}).  The operator $\Pi$ ensures that every
insertion of $V^{(2+1)}_{12,eff}$ represents scatterings coming from
different pairs of nucleons (i.e. the rungs of
Fig.~\ref{fig:resolvent30} alternate), preventing any double couting.
As there are no spectator nucleons in this expression, it is labeled
with the superscript $^{(3+0)}$.

Matrix elements $<H_{eff}>$ are given by
\begin{equation}\label{eqn:v3effME}
\begin{aligned}
\widetilde{<\Omega_i|}H_{eff}\widetilde{|\ \Omega_f>}
                       =&\\
\widetilde{<\Omega_i|}T_{eff}+\sum_{l<m}^3 &\left(V^{(2+1)}_{lm,eff}+V^{(3+0)}_{lm,eff}\right)
\widetilde{|\
                       \Omega_f>}\\
=\widetilde{<\Omega_i|}T_{eff}+3&\left(V^{(2+1)}_{12,eff}+V^{(3+0)}_{12,eff}\right)\widetilde{|\
                       \Omega_f>},
\end{aligned}
\end{equation}
where the factor of $3$ in the last expression comes from the
anti-symmetry of the states.  An expansion of Eq.~\ref{eqn:v3effME}
can now be found by expanding the resolvent of Eq.~\ref{eqn:V30eff} in
powers of $\Pi V^{(2+1)}_{12,eff}$,
\begin{equation*}
\begin{aligned}
\frac{1}{E-QT-Q\Pi
V^{(2+1)}_{12,eff}}Q&=\\
\frac{1}{E-QT}Q 
+\frac{1}{E-QT}&Q\Pi
V^{(2+1)}_{12,eff}\frac{1}{E-QT}Q+\ldots.
\end{aligned}
\end{equation*}
Expressions analogous to those of Eq.~(\ref{eqn:Gamma_n}) can
be constructed from  $\Pi$ and $V^{(2+1)}_{12,eff}$,
\begin{equation}\label{eqn:Gamma_ns}
\begin{aligned}
\Gamma_n=&PG_0\left(V^{(2+1)}_{12,eff}G_0\right)^nP\\
\gamma_n=&PG_0\left(\Pi V^{(2+1)}_{12,eff}G_0\right)^nP\\
\tilde{\Gamma}_n=&PG_0V^{(2+1)}_{12,eff}G_0\left(\Pi
V^{(2+1)}_{12,eff}G_0\right)^{n-1}P.
\end{aligned}
\end{equation}
Combining Eq.~\ref{eqn:Gamma_ns} with the expansion of
Eq.~\ref{eqn:v3effME} gives the final result (compare with
Eq.~\ref{eqn:bhfinal}),
\begin{equation}\label{eqn:bhfinal3}
\begin{aligned}
H_{eff}=&E-\Gamma_0^{-1}+3\Gamma_0^{-1}\Gamma_1\Gamma_0^{-1}+\\
3\Gamma_0^{-1}&\left\{\tilde{\Gamma}_2-\Gamma_1\Gamma_0^{-1}\gamma_1\right\}\Gamma_0^{-1}+\\
3\Gamma_0^{-1}&\{\tilde{\Gamma}_3-\tilde{\Gamma}_2\Gamma_0^{-1}\gamma_1-\Gamma_1\Gamma_0^{-1}\gamma_2+\\
&\quad\quad\quad\quad\quad\quad\Gamma_1\Gamma_0^{-1}\gamma_1\Gamma_0^{-1}\gamma_1\}\Gamma_0^{-1}+\ldots
\end{aligned}
\end{equation}
Figure~\ref{fig:threebodyenergies} shows the convergence of the
binding energies for $^3$H and $^3$He for several included-space
and oscillator parameter choices.  The expansion of
Eq.~(\ref{eqn:bhfinal3}) converges even for very small included
spaces, including $\Lambda=0$ (which corresponds to a single matrix
element).  Surprisingly, the most rapid convergence occurs for the
smallest included spaces.  We do not have a convincing physical
explanation for this result.

Note that the zeroth order results (i.e. $
E-\Gamma_0^{-1}+3\Gamma_0^{-1}\Gamma_1\Gamma_0^{-1}$) all overbind in
Fig.~\ref{fig:threebodyenergies}. This is possible since the zeroth
order calculation is no longer variational, as
$V^{(2+1)}_{12,eff}$ now depends on $E$ (see Eq.~\ref{eqn:V12eff}).
Hence there is no prescribed method for finding the optimal $b$, other
than by trial and error.  (We stress that fully converged results will
be independent of $b$, as we have executed the effective theory
faithfully.  By an optimal $b$ we mean one that will speed the
convergence.)  The values of $b$ used in
Fig.~\ref{fig:threebodyenergies} should be near the optimal.  Note
that these values are much larger than those found for the deuteron
calculations of Ref.\cite{Haxton:2002kb}.  Since $V^{(2+1)}_{12,eff}$
is calculated exactly, all short range two-body correlations are
correctly taken into account.  Hence $b$ is no longer forced to small
values by the demands of short-$r$ physics, but instead can relax to
values characteristic of the size of the three-body system.

As mentioned earlier, $V^{(2+1)}_{12,eff}$ is reminiscent of the
$G$-matrix found in traditional nuclear many-body theory, which is
also an infinite ladder sum in particle-particle propagation.  However
the differences are substantial.  Traditional shell-model calculations
have always used G-matrix interactions at the two-body level, ignoring
any dependence the operator may have on spectator nucleons (including
in general certain violations of the Pauli principle). The operator
$V^{(2+1)}_{12,eff}$, on the other hand, depends explicitly on the
spectator nucleon kinetic energy, as well as on the kinetic energies
of the interacting nucleons.  This dependence is manifest in
Eq.~\ref{eqn:V12eff}, as the operator $T$ in the resolvent
$\frac{E}{E-QT}Q$ represents the sum of the kinetic energies of all
nucleons.  This dependence on the spectator nucleon is essential for
calculating the correct two-body correlations within the three-body
system, as $Q$ is defined by the quanta carried by the three-body
configurations.  Similarly, for $A$-body calculations, the analogous
operator, $V^{(2+(A-2))}_{12,eff}$, would carry the dependence of all
spectator nucleons.

It appears that a perturbative expansion for the binding energy of
$^3$He/$^3$H results only if one first sums two-body ladder
interactions to all orders (\emph{i.e.}  $V^{(2+1)}_{12,eff}$).  This
summation softens the two-body interaction, which on iteration can
then generate the effective three-body interaction (\emph{i.e.}
$V^{(3+0)}_{12,eff}$) perturbatively.  A similar procedure was followed
in Refs.\cite{Bedaque:1999ve,Gabbiani:1999yv}, where the triton
binding energy was calculated by perturbing in di-baryon fields.  The
di-baryon fields themselves represent an infinite scattering of
two-body interactions, in analogy with $V^{(2+1)}_{12,eff}$.

\section{A non-perturbative three-body calculation\label{sect:nonperturbative}}

A drawback of the perturbative approach of the previous section is
that for each included space defined by $\Lambda$, there is only a
limited range of $b$s for which the expansion converges readily.
Furthermore, there is no definite procedure for estimating the optimal
$b$, as the zeroth-order calculation is no longer variational.  Thus a
nonperturbative procedure would be attractive: any $b$ could be
chosen, and physical observables would prove to be independent of $b$,
as they must for a rigorously executed effective theory.  Such a
non-perturbative summation is indeed possible by reshuffling the terms
of Eq.~(\ref{eqn:bhfinal3}) to form a summable geometric series and by
numerically solving an integral equation.  This approach provides an
opportunity to directly compare the relative sizes of
$V^{(2+1)}_{12,eff}$ and $V^{(3+0)}_{12,eff}$ as functions of both
$\Lambda$ and $b$.

In this section we only show results for the triton system, as $^3$He
calculations produce similar results\footnote{For $^3$He results, the
  reader is referred to Ref.\cite{Luu:2003}}.  The starting point this
time is Eq.~\ref{eqn:bhfinal3},
\begin{equation}\label{eqn:bhfinal3repeat}
\begin{aligned}
&H_{eff}=E-\Gamma_0^{-1}+3\Gamma_0^{-1}\Gamma_1\Gamma_0^{-1}+\\
&3\Gamma_0^{-1}\left\{\tilde{\Gamma}_2-\Gamma_1\Gamma_0^{-1}\gamma_1\right\}\Gamma_0^{-1}+\\
&3\Gamma_0^{-1}\left\{\tilde{\Gamma}_3-\tilde{\Gamma}_2\Gamma_0^{-1}\gamma_1-\Gamma_1\Gamma_0^{-1}
  \gamma_2+
\Gamma_1\Gamma_0^{-1}\gamma_1\Gamma_0^{-1}\gamma_1\right\}\Gamma_0^{-1}\\
&+\ldots,
\end{aligned}
\end{equation}
where the set $\{\Gamma_n,\tilde{\Gamma}_n,\gamma_n\}$ is given by
Eq.~\ref{eqn:Gamma_ns}.  This expansion can be rearranged into
\begin{widetext}
\begin{equation}
\begin{aligned}
H_{eff}&=E-\Gamma_0^{-1}+3\Gamma_0^{-1}\Gamma_1\Gamma_0^{-1}\\
       +&3(\Gamma_0^{-1}\left(\tilde{\Gamma}_2+\tilde{\Gamma}_3+\ldots\right)\Gamma_0^{-1}
       -\Gamma_0^{-1}\left(\tilde{\Gamma}_2+\tilde{\Gamma}_3+\ldots\right)\Gamma_0^{-1}
       \left(\gamma_1+\gamma_2+\ldots\right)\Gamma_0^{-1}+\ldots)\\
       -3(\Gamma_0^{-1}&\Gamma_1\Gamma_0^{-1}\left(\gamma_1+\gamma_2+\ldots\right)\Gamma_0^{-1}
       -\Gamma_0^{-1}\Gamma_1\Gamma_0^{-1}\left(\gamma_1+\gamma_2+\ldots\right)\Gamma_0^{-1}
       \left(\gamma_1+\gamma_2+\ldots\right)\Gamma_0^{-1}+\ldots).
\end{aligned}\label{eqn:3bhfinalnp}
\end{equation}
\end{widetext}
Equation~\ref{eqn:3bhfinalnp} can be verified by expanding its terms
out and directly comparing with Eq.~\ref{eqn:bhfinal3repeat}.  The
following definitions can be made to simplify
Eq.~\ref{eqn:3bhfinalnp},
\begin{equation}
\begin{aligned}
\tilde{\Gamma}_{\infty}&=\tilde{\Gamma}_2+\tilde{\Gamma}_3+\ldots\\
\gamma_{\infty}&=\gamma_1+\gamma_2+\ldots,
\end{aligned}\label{eqn:gamma_inftys}
\end{equation}
where the above expressions represent infinite summations.
Substituting these expressions into Eq.~\ref{eqn:3bhfinalnp} gives
\begin{equation}
\begin{aligned}
&H_{eff}=E-\Gamma_0^{-1}+3\Gamma_0^{-1}\Gamma_1\Gamma_0^{-1}\\
       +&3(\Gamma_0^{-1}\tilde{\Gamma}_{\infty}\Gamma_0^{-1}
       -\Gamma_0^{-1}\tilde{\Gamma}_{\infty}\Gamma_0^{-1}\gamma_{\infty}\Gamma_0^{-1}+\ldots)\\
       -&3(\Gamma_0^{-1}\Gamma_1\Gamma_0^{-1}\gamma_{\infty}\Gamma_0^{-1}
       -\Gamma_0^{-1}\Gamma_1\Gamma_0^{-1}\gamma_{\infty}\Gamma_0^{-1}\gamma_{\infty}\Gamma_0^{-1}+\ldots)
\end{aligned}\label{eqn:3bhfinalnp2}
\end{equation}
The two geometric series in the equation above can be summed to give
the final desired result,
\begin{equation}\label{eqn:3bodynpert}
\begin{aligned}
H_{eff}=&E-\Gamma_0^{-1}+3\Gamma_0^{-1}\Gamma_1\Gamma_0^{-1}\\
       +&3\Gamma_0^{-1}\left(\tilde{\Gamma}_{\infty}-\Gamma_1\Gamma_0^{-1}\gamma_{\infty}\right)
       \frac{1}{1+\Gamma_0^{-1}\gamma_{\infty}}\Gamma_0^{-1}\\
       =&T_{eff}+3V^{(2+1)}_{12,eff}+3V^{(3+0)}_{12,eff},
\end{aligned}
\end{equation}
where we have made the following identifications,
\begin{displaymath}
\begin{aligned}
T_{eff}=&E-\Gamma_0^{-1}\\
V^{(2+1)}_{12,eff}=&\Gamma_0^{-1}\Gamma_1\Gamma_0^{-1}\\
V^{(3+0)}_{12,eff}=&\Gamma_0^{-1}\left(\tilde{\Gamma}_{\infty}-\Gamma_1\Gamma_0^{-1}\gamma_{\infty}\right)
\frac{1}{1+\Gamma_0^{-1}\gamma_{\infty}}\Gamma_0^{-1}.
\end{aligned}
\end{displaymath}

The usefulness of Eq.~(\ref{eqn:3bodynpert}) depends on having an
efficient method for calculating the included-space matrix elements of
$\gamma_{\infty}$ and $\tilde{\Gamma}_{\infty}$.  This can be done by
first considering the state $\gamma_{\infty}|\ 
\Omega_{\sigma}>=|\Psi_{\sigma}>$, which, using
Eq.~\ref{eqn:gamma_inftys}, satisfies
\begin{equation}\label{eqn:3bodyintegraleq}
\begin{aligned}
|\Psi_{\sigma}>=&\gamma_1|\ \Omega_{\sigma}>+\gamma_2|\ \Omega_{\sigma}>+\ldots\\
=G_0&\Pi V_{12}G_0|\ \Omega_{\sigma}>+G_0\Pi
V_{12}G_0\Pi V_{12}G_0|\ \Omega_{\sigma}>+\ldots\\ =
G_0&\Pi
V_{12}G_0|\ \Omega_{\sigma}>+G_0\Pi V_{12}|\Psi_{\sigma}>.
\end{aligned}
\end{equation}
Hence $|\Psi_{\sigma}>$ satisfies the Fredholm integral equation of
the second kind\cite{Hackbusch:1995}.  There are numerous numerical
methods available for solving this particular integral equation.  We
use a general projection algorithm\cite{Hackbusch:1995}.  Due to the
large number of partial waves and the complicated structure of the
operator $\Pi$, solving the integral equation by matrix inversion is 
impractical.

Once $|\Psi_{\sigma}>$ is found, the matrix elements
\begin{displaymath}
<\Omega_{i}|\gamma_{\infty}|\
\Omega_{\sigma}>=<\Omega_{i}|\Psi_{\sigma}>
\end{displaymath}
can be evaluated. Then the matrix elements
$<\Omega_{i}|\tilde{\Gamma}_{\infty}|\Omega_{\sigma}>$ are easily
calculated since
\begin{equation*}
\tilde{\Gamma}_{\infty}|\Omega_{\sigma}>=G_0V_{12}|\Psi_{\sigma}>,
\end{equation*}
which can be verified using Eqs.~\ref{eqn:Gamma_ns}
and~\ref{eqn:gamma_inftys}.  Hence
\begin{displaymath}
<\Omega_{i}|\tilde{\Gamma}_{\infty}|\
\Omega_{\sigma}>=<\Omega_{i}|G_0V_{12}|\Psi_{\sigma}>.
\end{displaymath}

Tables~\ref{tab:triton_results} and~\ref{tab:tritonNP3half} show the
calculated binding energies for the triton system as a function of the
included-space size $\Lambda$ and oscillator parameter $b$.  In
Table~\ref{tab:triton_results} we have ignored the small
isospin-violating parts of the Av18 potential.  Hence isospin is
conserved at $T=1/2$.  Table~\ref{tab:tritonNP3half} includes
isospin-violating contributions.  This causes a small admixture of
$T=3/2$ components into the ground state. At about the keV level, the
results are independent of the choice of the included space, i.e., of
$b$ and $\Lambda$.  The $\sim$ few keV variations illustrate the level
of our numerical precision.  Note that for the values $b$=.83 and 1.17
fm, our calculations of the previous section would give non-converging
binding energies.  This of course emphasizes the non-perturbative
nature of these calculations.

Note that Eq.~\ref{eqn:3bodynpert} has the correct limiting behavior
as $\Lambda \rightarrow \infty$, $H_{eff} \rightarrow H$.  In this
limit the projection operator $P\rightarrow 1$, giving
$\Gamma_0^{-1} \rightarrow G_0^{-1}$. Thus
\begin{displaymath}
E-\Gamma_0^{-1}\overset{\Lambda\rightarrow\infty}{\longrightarrow} T.
\end{displaymath}
It is also straightforward to show that
\begin{displaymath}
\begin{aligned}
V^{(2+1)}_{12,eff}=\Gamma_0^{-1}\Gamma_1\Gamma_0^{-1}&\overset{\Lambda\rightarrow\infty}{\longrightarrow}
\frac{1}{1+t_{12}G_0}t_{12}\\
&=V_{12},
\end{aligned}
\end{displaymath}
where we have used Eq.~\ref{eqn:t12def} to obtain the second line
above.  Finally, it is obvious that $V^{(3+0)}_{12,eff}$ vanishes in
this limit since 
\begin{displaymath}
\tilde{\Gamma}_{\infty}-\Gamma_1\Gamma_0^{-1}\gamma_{\infty}\overset{\Lambda\rightarrow\infty}
{\longrightarrow}
\quad\tilde{\Gamma}_{\infty}-G_0V\gamma_{\infty}=0.
\end{displaymath}
Hence $H_{eff}\rightarrow H$.

Tables~\ref{tab:00tritonME}-~\ref{tab:22tritonME} show the variation
of the matrix elements of $V^{(2+1)}_{12,eff}$ and
$V^{(3+0)}_{12,eff}$ with $\Lambda$ and $b$.  Isospin-violating
effects have also been ignored in these results.  As the number of
included-space examples is small, it is difficult to extract any
limiting behavior.  Yet it does seem that as $\Lambda$ increases,
matrix elements of $V^{(2+1)}_{12,eff}$ approach the bare $V_{12}$
(indicated by the symbol $\infty$ in the tables).  This is especially
evident in Table~\ref{tab:02tritonME}, where the effect of
renormalization is small (i.e. the renormalized matrix elements are
not too different from their bare matrix elements).  In the cases
involving $V^{(3+0)}_{12,eff}$, limiting behavior is also difficult to
extract.  Indeed, in Tables.~\ref{tab:00tritonME}
and~\ref{tab:22tritonME} matrix elements of $V^{(3+0)}_{12,eff}$ tend
to grow away from zero with increasing $\Lambda$ rather than tend
toward zero.  However, in these cases the renormalization is strong,
so that limiting behavior would not be expected for such small
$\Lambda$.  In Table~\ref{tab:02tritonME} the renormalization is much
weaker, and the limiting behavior $3V^{(3+0)}_{12,eff}\rightarrow 0$
is clearly seen for the $\hbar\omega=15$ and $=60$ MeV examples.  It
is also clear from Tables.~\ref{tab:00tritonME}-~\ref{tab:22tritonME}
that the matrix elements of $V^{(2+1)}_{12,eff}$ and
$V^{(3+0)}_{12,eff}$ have a strong dependence on $\Lambda$ and $b$
($\hbar\omega$), a dependence often ignored in traditional shell-model
calculations.

Generally the contribution of $<V^{(3+0)}_{12,eff}>$ to the overall
binding energy is small compared to $<V^{(2+1)}_{12,eff}>$ ($\sim 10\%
$ of $V^{(2+1)}_{12,eff}$), as is evident from
Fig.~\ref{fig:V21_vs_V30}. This graph gives the $<T_{eff}>$,
$3<V^{(2+1)}_{12,eff}>$, and $3<V^{(3+0)}_{12,eff}>$ contributions to
the binding energy as functions of $\Lambda$ and $b$.  Interestingly,
for each included-space $\Lambda$, there are two values of $b$ at
which the contribution of $<V^{(3+0)}_{12,eff}>$ completely vanishes.
Table~\ref{tab:b_zeros} lists specific values.  If this result proves
to be a generic property of the three-body $H_{eff}$ in heavier nuclei
(that is, if zeros appear in those calculations at approximately the
same value of $b$), this could greatly simplify the more complicated
included-space diagonalizations required in heavy systems.  For
example, the three-body contribution could be ignored or could be
explored in low-order perturbation theory.  This would require
establishing that the A=3 zero for $<V^{(3+0)}_{12,eff}>$ does persist
at about the same $b$ in heavier systems.  This seems plausible, as
the three-body ladder in a heavy nucleus is quite similar to that in
A=3.

\section{Conclusion\label{sect:conclusion}}

\begin{figure*}
\begin{center}
\begingroup%
  \makeatletter%
  \newcommand{\GNUPLOTspecial}{%
    \@sanitize\catcode`\%=14\relax\special}%
  \setlength{\unitlength}{0.1bp}%
\begin{picture}(3600,3023)(0,0)%
\includegraphics{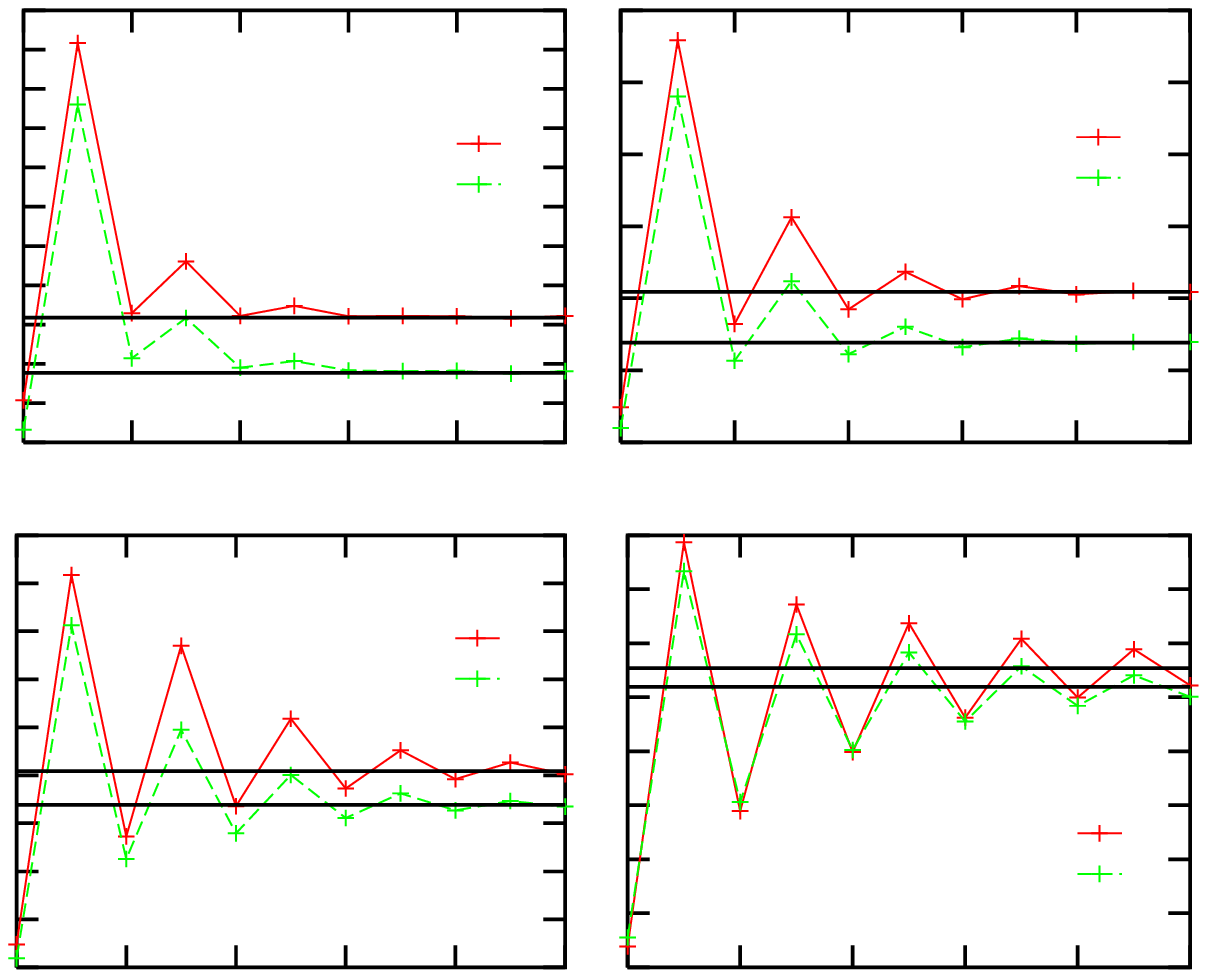}
\fontsize{8}{\baselineskip}\selectfont
\put(3196,389){\makebox(0,0)[r]{Triton}}%
\put(3196,506){\makebox(0,0)[r]{Helium3}}%
\put(3196,584){\makebox(0,0)[r]{ b=1.66 fm}}%
\put(3196,664){\makebox(0,0)[r]{$\Lambda\hbar\omega=6\hbar\omega$}}%
\put(2730,20){\makebox(0,0){(d)}}%
\put(3540,80){\makebox(0,0){ 10}}%
\put(3216,80){\makebox(0,0){ 8}}%
\put(2892,80){\makebox(0,0){ 6}}%
\put(2568,80){\makebox(0,0){ 4}}%
\put(2244,80){\makebox(0,0){ 2}}%
\put(1920,80){\makebox(0,0){ 0}}%
\put(1900,1364){\makebox(0,0)[r]{-2}}%
\put(1900,1209){\makebox(0,0)[r]{-4}}%
\put(1900,1053){\makebox(0,0)[r]{-6}}%
\put(1900,898){\makebox(0,0)[r]{-8}}%
\put(1900,742){\makebox(0,0)[r]{-10}}%
\put(1900,587){\makebox(0,0)[r]{-12}}%
\put(1900,431){\makebox(0,0)[r]{-14}}%
\put(1900,276){\makebox(0,0)[r]{-16}}%
\put(1900,120){\makebox(0,0)[r]{-18}}%
\put(3192,2394){\makebox(0,0)[r]{Triton}}%
\put(3192,2511){\makebox(0,0)[r]{Helium3}}%
\put(3192,2589){\makebox(0,0)[r]{ b=2.04 fm}}%
\put(3192,2669){\makebox(0,0)[r]{$\Lambda\hbar\omega=2\hbar\omega$}}%
\put(2720,1532){\makebox(0,0){(b)}}%
\put(3540,1592){\makebox(0,0){ 10}}%
\put(3212,1592){\makebox(0,0){ 8}}%
\put(2884,1592){\makebox(0,0){ 6}}%
\put(2556,1592){\makebox(0,0){ 4}}%
\put(2228,1592){\makebox(0,0){ 2}}%
\put(1900,1592){\makebox(0,0){ 0}}%
\put(1880,2876){\makebox(0,0)[r]{-3}}%
\put(1880,2669){\makebox(0,0)[r]{-4}}%
\put(1880,2461){\makebox(0,0)[r]{-5}}%
\put(1880,2254){\makebox(0,0)[r]{-6}}%
\put(1880,2047){\makebox(0,0)[r]{-7}}%
\put(1880,1839){\makebox(0,0)[r]{-8}}%
\put(1880,1632){\makebox(0,0)[r]{-9}}%
\put(1404,951){\makebox(0,0)[r]{Triton}}%
\put(1404,1068){\makebox(0,0)[r]{Helium3}}%
\put(1404,1146){\makebox(0,0)[r]{ b=2.04 fm}}%
\put(1404,1226){\makebox(0,0)[r]{$\Lambda\hbar\omega=4\hbar\omega$}}%
\put(950,20){\makebox(0,0){(c)}}%
\put(40,742){%
\makebox(0,0)[b]{\shortstack{MeV}}%
}%
\put(1740,80){\makebox(0,0){ 10}}%
\put(1424,80){\makebox(0,0){ 8}}%
\put(1108,80){\makebox(0,0){ 6}}%
\put(792,80){\makebox(0,0){ 4}}%
\put(476,80){\makebox(0,0){ 2}}%
\put(160,80){\makebox(0,0){ 0}}%
\put(140,1364){\makebox(0,0)[r]{-2}}%
\put(140,1226){\makebox(0,0)[r]{-3}}%
\put(140,1088){\makebox(0,0)[r]{-4}}%
\put(140,949){\makebox(0,0)[r]{-5}}%
\put(140,811){\makebox(0,0)[r]{-6}}%
\put(140,673){\makebox(0,0)[r]{-7}}%
\put(140,535){\makebox(0,0)[r]{-8}}%
\put(140,396){\makebox(0,0)[r]{-9}}%
\put(140,258){\makebox(0,0)[r]{-10}}%
\put(140,120){\makebox(0,0)[r]{-11}}%
\put(1408,2375){\makebox(0,0)[r]{Triton}}%
\put(1408,2492){\makebox(0,0)[r]{Helium3}}%
\put(1408,2570){\makebox(0,0)[r]{ b=1.66 fm}}%
\put(1408,2650){\makebox(0,0)[r]{$\Lambda\hbar\omega=0\hbar\omega$}}%
\put(960,1532){\makebox(0,0){(a)}}%
\put(40,2254){%
\makebox(0,0)[b]{\shortstack{MeV}}%
}%
\put(1740,1592){\makebox(0,0){ 10}}%
\put(1428,1592){\makebox(0,0){ 8}}%
\put(1116,1592){\makebox(0,0){ 6}}%
\put(804,1592){\makebox(0,0){ 4}}%
\put(492,1592){\makebox(0,0){ 2}}%
\put(180,1592){\makebox(0,0){ 0}}%
\put(160,2876){\makebox(0,0)[r]{-3}}%
\put(160,2763){\makebox(0,0)[r]{-3.5}}%
\put(160,2650){\makebox(0,0)[r]{-4}}%
\put(160,2537){\makebox(0,0)[r]{-4.5}}%
\put(160,2424){\makebox(0,0)[r]{-5}}%
\put(160,2311){\makebox(0,0)[r]{-5.5}}%
\put(160,2197){\makebox(0,0)[r]{-6}}%
\put(160,2084){\makebox(0,0)[r]{-6.5}}%
\put(160,1971){\makebox(0,0)[r]{-7}}%
\put(160,1858){\makebox(0,0)[r]{-7.5}}%
\put(160,1745){\makebox(0,0)[r]{-8}}%
\put(160,1632){\makebox(0,0)[r]{-8.5}}%
\end{picture}%
\endgroup
\caption[Perturbative calculation of triton and $^3$He B.Es.]
{Perturbative calculation of triton and $^3$He B.Es. for different 
$\Lambda$s and $b$s using the expansion given in
Eq.~\ref{eqn:bhfinal3}.  The solid black lines of each graph
correspond to the exact binding energies. \label{fig:threebodyenergies}}
\end{center}
\end{figure*}

\begin{figure}
\begin{center}
\begin{picture}(0,0)%
\includegraphics{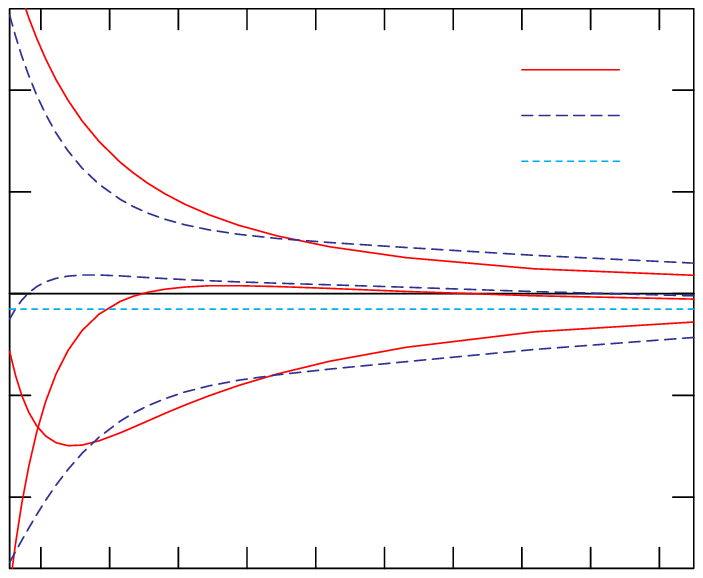}%
\end{picture}%
\setlength{\unitlength}{0.0200bp}%
\begin{picture}(12599,10259)(0,0)%
\put(1650,2676){\makebox(0,0)[r]{\strut{}-100}}%
\put(1650,4141){\makebox(0,0)[r]{\strut{}-50}}%
\put(1650,5607){\makebox(0,0)[r]{\strut{} 0}}%
\put(1650,7072){\makebox(0,0)[r]{\strut{} 50}}%
\put(1650,8538){\makebox(0,0)[r]{\strut{} 100}}%
\put(2375,1100){\makebox(0,0){\strut{} 0.6}}%
\put(3364,1100){\makebox(0,0){\strut{} 0.8}}%
\put(4354,1100){\makebox(0,0){\strut{} 1}}%
\put(5343,1100){\makebox(0,0){\strut{} 1.2}}%
\put(6333,1100){\makebox(0,0){\strut{} 1.4}}%
\put(7322,1100){\makebox(0,0){\strut{} 1.6}}%
\put(8312,1100){\makebox(0,0){\strut{} 1.8}}%
\put(9301,1100){\makebox(0,0){\strut{} 2}}%
\put(10291,1100){\makebox(0,0){\strut{} 2.2}}%
\put(11280,1100){\makebox(0,0){\strut{} 2.4}}%
\put(1100,5680){\rotatebox{90}{\makebox(0,0){\strut{}MeV}}}%
\put(6850,275){\makebox(0,0){\strut{}Oscillator b}}%
\put(4354,7365){\makebox(0,0)[l]{\strut{}$T_{eff}$}}%
\put(4354,5167){\makebox(0,0)[l]{\strut{}$3V^{(3+0)}_{eff}$}}%
\put(4354,3702){\makebox(0,0)[l]{\strut{}$3V^{(2+1)}_{eff}$}}%
\put(4354,7365){\makebox(0,0)[l]{\strut{}$T_{eff}$}}%
\put(4354,5167){\makebox(0,0)[l]{\strut{}$3V^{(3+0)}_{eff}$}}%
\put(4354,3702){\makebox(0,0)[l]{\strut{}$3V^{(2+1)}_{eff}$}}%
\put(4354,7365){\makebox(0,0)[l]{\strut{}$T_{eff}$}}%
\put(4354,5167){\makebox(0,0)[l]{\strut{}$3V^{(3+0)}_{eff}$}}%
\put(4354,3702){\makebox(0,0)[l]{\strut{}$3V^{(2+1)}_{eff}$}}%
\put(4354,7365){\makebox(0,0)[l]{\strut{}$T_{eff}$}}%
\put(4354,5167){\makebox(0,0)[l]{\strut{}$3V^{(3+0)}_{eff}$}}%
\put(4354,3702){\makebox(0,0)[l]{\strut{}$3V^{(2+1)}_{eff}$}}%
\put(4354,7365){\makebox(0,0)[l]{\strut{}$T_{eff}$}}%
\put(4354,5167){\makebox(0,0)[l]{\strut{}$3V^{(3+0)}_{eff}$}}%
\put(4354,3702){\makebox(0,0)[l]{\strut{}$3V^{(2+1)}_{eff}$}}%
\put(4354,7365){\makebox(0,0)[l]{\strut{}$T_{eff}$}}%
\put(4354,5167){\makebox(0,0)[l]{\strut{}$3V^{(3+0)}_{eff}$}}%
\put(4354,3702){\makebox(0,0)[l]{\strut{}$3V^{(2+1)}_{eff}$}}%
\put(9026,8831){\makebox(0,0)[r]{\strut{}$0\hbar\omega$}}%
\put(9026,8172){\makebox(0,0)[r]{\strut{}$2\hbar\omega$}}%
\put(9026,7513){\makebox(0,0)[r]{\strut{}Sum}}%
\end{picture}%
\caption[Relative sizes of $3V^{(2+1)}_{12,eff}$ and
$3V^{(3+0)}_{12,eff}$.]
{Relative sizes of $T_{eff}$, $3V^{(2+1)}_{12,eff}$, and 
$3V^{(3+0)}_{12,eff}$ for $0\hbar\omega$ and $2\hbar\omega$ triton calculations 
as a function of $b$.  The upper pairs of lines refer to $T_{eff}$,
the middle pair $V^{(3+0)}_{12,eff}$,and the bottom two
$V^{(2+1)}_{12,eff}$.  Note that the respective sums of all three
terms gives the triton binding energy.}
\label{fig:V21_vs_V30}
\end{center}
\end{figure}

We found in Section~\ref{sect:perturbative} that the simple
perturbative scheme found in Ref.\cite{Haxton:2002kb} (\emph{i.e.}
Eq.~(\ref{eqn:bhfinal})) did not directly extend to the three-body
system due to the additional length scale introduced by the third
nucleon.  For any choice of Jacobi coordinates, this second length
scale excludes a class of hard-core interactions from the included
space regardless of the choice of $b$.  This problem was circumvented
by invoking the Faddeev decomposition on $H_{eff}$, which allowed us
to sum to all orders the repeated potential scattering between the
same two nucleons, generating $V^{(2+1)}_{12,eff}$.  Hence the strong
repulsive two-nucleon force was treated non-perturbatively.  The
remaining contribution to $H_{eff}$, $V^{(3+0)}_{12,eff}$, can then be
calculated by summing repeated insertions of $V^{(2+1)}_{12,eff}$ on
alternate pairs of nucleons.  For certain ranges of $b$, this
expansion converged after several orders of perturbation even for the
smallest allowed included spaces, as shown in
Fig.~\ref{fig:threebodyenergies}. 

In Section~\ref{sect:nonperturbative} we described a non-perturbative
method for calculating $H_{eff}$ in the three-body system, summarized
in Eq.~(\ref{eqn:3bodynpert}).  This allowed a direct comparison of
the contributions of $<T_{eff}>$, $<V^{(2+1)}_{12,eff}>$, and
$<V^{(3+0)}_{12,eff}>$.  Because the calculation was non-perturbative,
we were able to explore the dependence of the included-space matrix
elements (and establish the independence of the binding energy) for a
wider range of $b$s and $\Lambda$s.  Also, for each included space, we
found two values of $b$ where the total binding was due to
$V^{(2+1)}_{12,eff}$, a result that should be explored in heavier
systems.  The numerical methods included an efficient algorithm for
solving integral equations (\emph{i.e.}
Eq.~(\ref{eqn:3bodyintegraleq})).  In the current applications to A=3
systems, the numerical effort was not substantial.

Work is in progress to apply a similar formalism to the alpha
particle.  As in the three-body case, one must account for the various
length scales inherent to the four-body system.  This can be done by
invoking the Faddeev-Yakubovsky decompositions on $H_{eff}$, in direct
analogy to the three-body decompositions presented in this paper.
Such a procedure will not only yield $<T_{eff}>$,
$<V^{(2+2)}_{12,eff}>$ and $<V^{(3+1)}_{12,eff}>$ (direct analogs of
$<V^{(2+1)}_{12,eff}>$ and $<V^{(3+0)}_{12,eff}>$), but also
$<V^{(4+0)}_{12,eff}>$.  This last expression represents the induced
effective four-body interaction.  It will be interesting to see if
these calculations verify the commonly implicit assumption of the
hierarchical sizes of these interactions (\emph{i.e.}
$V^{(2+2)}_{12,eff}\ <\ V^{(3+1)}_{12,eff}\ <\ V^{(4+0)}_{12,eff}$).
The four-body effective interaction is the most complicated
lowest-order (s-wave) operator that can be constructed. Furthermore,
alpha-particle clustering is an important phenomenological feature of
nuclear structure, apparent even in simple systems like $^8$Be.  Thus
it will be interesting to determine the size of this contribution. The
A=4 calculations will also check whether zeros in the three-body
effective interaction persist.  If they do, we may learn something
about their ``trajectories'' in $b$ as A is inceased.

\begin{widetext}
\appendix

\section{Derivation of Eq.~\ref{eqn:bhfinal}\label{app:bhfinal}}
For completeness, we show the derivation of Eq.~\ref{eqn:BHre},
which was first used in Ref.\cite{Haxton:2002kb} in deriving a
perturbative expansion for the deuteron.
We begin by explicitly expressing the BH equation in its various components,
\begin{equation}\label{eqn:heffexpand}
\begin{aligned}
H_{eff}=&H+H\frac{1}{E-QH}QH\\
       =&H+T\frac{1}{E-QH}QT+V\frac{1}{E-QH}QV+T\frac{1}{E-QH}QV+V\frac{1}{E-QH}QT.
\end{aligned}
\end{equation}
Next consider the operator
\begin{equation}\label{eqn:teffexpand}
\begin{aligned}
T\frac{1}{E-QH}QT=&T\left[\frac{1}{E-QT}+\frac{1}{E-QT}QV\frac{1}{E-QH}\right]QT\\
       =&T\frac{1}{E-QT}\left[1+QV\frac{1}{E-QH}\right]QT\\
       =&T\frac{1}{E-QT}\left[1+QV\left\{\frac{1}{E-QT}+\frac{1}{E-QH}QV\frac{1}{E-QT}\right\}\right]QT\\
       =&TQ\frac{1}{E-TQ}\left[E-QT+QV+QV\frac{1}{E-QH}QV\right]\frac{1}{E-QT}QT,
\end{aligned}
\end{equation}
where the relation $\frac{1}{E-QT}QT=TQ\frac{1}{E-TQ}$ was used.  A
similar calculation gives
\begin{equation}\label{eqn:vteffexpand}
\begin{aligned}
T\frac{1}{E-QH}QV=&T\frac{1}{E-QT}\left[V+QV\frac{1}{E-QH}QV\right],\\
V\frac{1}{E-QH}QT=&\left[V+V\frac{1}{E-QH}QV\right]\frac{1}{E-QT}QT.
\end{aligned}
\end{equation}
Substituting Eqs.~\ref{eqn:teffexpand} and~\ref{eqn:vteffexpand} into
Eq.~\ref{eqn:heffexpand} and using the fact that
\begin{displaymath}
1+\frac{1}{E-QT}QT=\frac{E}{E-QT},
\end{displaymath}
and
\begin{displaymath}
T+TQ\frac{1}{E-TQ}(E-T)\frac{1}{E-QT}QT=TQ\frac{1}{E-TQ}(T-T\frac{Q}{E}T)\frac{1}{E-QT}QT,
\end{displaymath}
gives the desired result:
\begin{displaymath}
H_{eff}=TQ\frac{1}{E-TQ}(T-T\frac{Q}{E}T)\frac{1}{E-QT}QT+TQ\frac{1}{E-TQ}(V+V\frac{1}{E-QH}QV)\frac{1}
{E-QT}QT.
\end{displaymath}
Hence
\begin{equation}
<n'\ l'\ \alpha|H_{eff}|n\ l\ \alpha>=\widetilde{<n'\ l'\
\alpha|}T_{eff}+V_{eff}\widetilde{|n\ l\ \alpha>},
\label{eqn:newbh}
\end{equation}
where
\begin{eqnarray}
T_{eff}&=&T+T\frac{-Q}{E}T \label{eqn:appTeff}\\
V_{eff}&=&V+V\frac{1}{E-QH}QV, \label{eqn:appVeff}
\end{eqnarray}
and
\begin{equation}
\widetilde{|n\ l\ \alpha>}=\frac{E}{E-QT}|n\ l\
\alpha>.\label{eqn:endshellstate}
\end{equation}
Note that the resolvent in Eq.~\ref{eqn:appVeff} is now sandwiched
between $V$s only, and not $T$s.  Forming a perturbative expansion
involves expanding this resolvent as was done in
Eq.~\ref{eqn:resolvent_expansion}.

The operator $\frac{E}{E-QT}$ acts non-trivially only on endshell
states.  That is,
\begin{equation}
\frac{E}{E-QT}|n\ l\ \alpha>=\left\{ \begin{array}{ll} \widetilde{|n\
                    l\ \alpha>} &\mbox{if $|n\ l\ \alpha>$ endshell}\\
                    |n\ l\ \alpha>              &\mbox{if $|n\ l\
                    \alpha>$ below endshell,}
                  \end{array}
                    \right.
\label{eqn:endshellrelation}
\end{equation}
which can be easily verified by expanding the operator in powers of
$QT$ and noting the fact that the kinetic energy operator $T$ can change the
principal quantum number $n$ by at most $\pm 1$.  These `tilde' states
$\widetilde{|n\ l\ \alpha>}$ represent the culmination of the
infinitely repeated scatterings of the operator $QT$ on the original
states $|n\ l\ \alpha>$.  Such scatterings persist at large-$r$ even
though the potential $V$ has already vanished.

\section{Derivation of $V_{12,eff}^{(2+1)}$ (Eq.~\ref{eqn:exactV12eff})}\label{sect:exactV12eff}

Iterating the last expression of Eq.~\ref{eqn:V12eff} on
$V_{12,eff}^{(2+1)}$ gives
\begin{equation}\label{eqn:V12effExpansion1}
\begin{aligned}
V_{12,eff}^{(2+1)}=&V_{12}+V_{12}\frac{1}{E-QT}QV_{12}+\\
                   &V_{12}\frac{1}{E-QT}QV_{12}\frac{1}{E-QT}QV_{12}+\ldots
\end{aligned}
\end{equation}
Replacing $\frac{1}{E-QT}Q$ by the relevant expression below Eq.~\ref{eqn:invGamma0},
Eq.~\ref{eqn:V12effExpansion1} becomes
\begin{equation}\label{eqn:V12effExpansion2}
\begin{aligned}
V_{12,eff}^{(2+1)}=&V_{12}+V_{12}\left(G_0-G_0\Gamma_0^{-1}G_0\right)V_{12}\\
                  +&V_{12}\left(G_0-G_0\Gamma_0^{-1}G_0\right)V_{12}\left(G_0-G_0\Gamma_0^{-1}G_0\right)
                  V_{12}\\
                  +&\ldots
\end{aligned}
\end{equation}
The various terms of Eq.~\ref{eqn:V12effExpansion2} can be re-ordered
to give
\begin{multline}\label{eqn:V12effExpansion3}
V_{12,eff}^{(2+1)}=\left(V_{12}+V_{12}G_0V_{12}+V_{12}G_0V_{12}G_0V_{12}+\ldots\right)\\
-\left(V_{12}+V_{12}G_0V_{12}+\ldots\right)G_0\Gamma_0^{-1}G_0\left(V_{12}+V_{12}G_0V_{12}+\ldots\right)\\
+\left(V_{12}+V_{12}G_0V_{12}\ldots\right)G_0\Gamma_0^{-1}G_0\left(V_{12}+V_{12}G_0V_{12}\ldots\right)G_0
\Gamma_0^{-1}G_0\left(V_{12}+V_{12}G_0V_{12}\ldots\right)-\ldots,
\end{multline}
where
\begin{equation}\label{eqn:V12sum}
V_{12}+V_{12}G_0V_{12}+V_{12}G_0V_{12}G_0V_{12}+\ldots
\end{equation}
represents an infinite sum of two-body interactions between particles
1 and 2.  The solution to Eq.~\ref{eqn:V12sum} is given by the operator $t_{12}$, where
\begin{equation}\label{eqn:t12}
t_{12}=V_{12}+V_{12}G_0t_{12}.
\end{equation}
As mentioned near the end of sect.~\ref{sect:perturbative}, the
operator $t_{12}$ differs from the usual two-body $t$-matrix since it
is imbedded within a three-body space.   Hence $G_0$ contains a dependence
on the Jacobi momentum $q$ of the third `spectator' nucleon.  This
dependence carries over to the operator $t_{12}$.  Hence the
designation $(2+1)$-body $t$-matrix.  

Replacing the infinite sum of Eq.~\ref{eqn:V12sum} by $t_{12}$ and
inserting this into Eq.~\ref{eqn:V12effExpansion3} gives
\begin{equation}\label{eqn:V12effExpansion4}
\begin{aligned}
V_{12,eff}^{(2+1)}=&t_{12}-t_{12}G_0\Gamma_0^{-1}G_0t_{12}\\
&+t_{12}G_0\Gamma_0^{-1}\Gamma_{\infty}^{(2+1)}\Gamma_0^{-1}G_0t_{12}\\
&-t_{12}G_0\Gamma_0^{-1}\Gamma_{\infty}^{(2+1)}\Gamma_0^{-1}\Gamma_{\infty}^{(2+1)}\Gamma_0^{-1}
G_0t_{12}\\
&+\ldots,
\end{aligned}
\end{equation}
where $\Gamma^{(2+1)}_{\infty}$ is given by Eq.~\ref{eqn:gamma21def},
i.e.
\begin{displaymath}
\Gamma^{(2+1)}_{\infty}=PG_0t_{12}G_0P.
\end{displaymath}
Equation~\ref{eqn:V12effExpansion4} can now be summed geometrically,
giving the desired result
\begin{displaymath}
V^{(2+1)}_{12,eff}=t_{12}-t_{12}G_0\left[\Gamma_0+\Gamma^{(2+1)}_{\infty}\right]^{-1}G_0t_{12}.
\end{displaymath}

\section{Matrix elements of $\frac{1}{E-T}$ for the three-body problem\label{sect:3bodyETME}}

The relevant matrix element is
\begin{equation}\label{eqn:3Mea}
<n'l';N'L'|\frac{1}{E-T}|nl;NL>= \delta_{l',l}\delta_{L',L}\int dp
dp'
R^*_{n',l}(p)R^*_{N',L}(p')\frac{p^2p'^2}{E-p^2/2m-p'^2/2m}R_{n,l}(p)R_{N,L}(p').
\end{equation}
Replacing the radial integrals by their series expansion and changing
to dimensionless variables, Eq.~\ref{eqn:3Mea} becomes
\begin{multline}\label{eqn:3Meb}
\delta_{l',l}\delta_{L',L}(-1)^{n+n'+N+N'}\sqrt{C_1(n,n',l;,N,N',L)}
\frac{2}{\hbar\omega}\\
\times \sum_{\substack{m,m'=0\\M,M'=0}}^{\substack{n,n'\\N,N'}}
\frac{(-1)^{m+m'+M+M'}}{\sqrt{C_2(n,n',l;N,N',L;m,m',M,M')}} \int dx
dx'
e^{-(x^2+x'^2)}\frac{x^{2(1+l+m+m')}x'^{2(1+L+M+M')}}{\frac{2E}{\hbar\omega}-x^2-x'^2},
\end{multline}
where
\begin{displaymath}
C_1(n,n',l;N,N',L)= 4\Gamma(n)\Gamma(n')\Gamma(N)\Gamma(N')
\Gamma(n+l+1/2)\Gamma(n'+l+1/2)\Gamma(N+L+1/2)\Gamma(N'+L+1/2),
\end{displaymath}
and
\begin{multline}
C_2(n,n',l;N,N',L;m,m',M,M')=\\
\Gamma(n-m)\Gamma(n'-m')\Gamma(N-M)\Gamma(N'-M')
\Gamma(l+m+3/2)\Gamma(l+m'+3/2)\\
\Gamma(L+M+3/2)\Gamma(L'+M'+3/2)\Gamma(m+1)\Gamma(m'+1)\Gamma(M+1)\Gamma(M'+1).
\end{multline}
An analytic solution to the integral in Eq.~\ref{eqn:3Meb} can be
found by first changing variables to cylindrical coordinates:
\begin{displaymath}
\begin{aligned}
x=&\rho cos\phi\\ x'=&\rho sin\phi\\ \rho^2=&x^2+x'^2.
\end{aligned}
\end{displaymath}
Hence the integral becomes
\begin{multline}\label{eqn:3Mec}
\int_0^{\infty} dx dx'
e^{-(x^2+x'^2)}\frac{x^{2(1+l+m+m')}x'^{2(1+L+M+M')}}{\frac{2E}{\hbar\omega}-x^2-x'^2}=
\left[\int_0^{\infty} d\rho\
e^{-\rho^2}\frac{\rho^{2(2+l+m+m'+L+M+M')+1}}{\frac{2E}{\hbar\omega}-\rho^2}\right]\\
\times
\left[\int_0^{\pi/2}d\phi \
(cos\phi)^{2(1+l+m+m')}(sin\phi)^{2(1+L+M+M')}\right].
\end{multline}
The integrals in square brackets have analytic solutions that can be
found in any book of integrals (e.g. see Ref.\cite{Gradshteyn:1965}).
They are written here for completeness:
\begin{multline}\label{eqn:integral1}
\int_0^{\infty} d\rho\
e^{-\rho^2}\frac{\rho^{2(2+l+m+m'+L+M+M')+1}}{\frac{2E}{\hbar\omega}-\rho^2}=
\frac{(-1)^{3+l+m+m'+L+M+M'+1}}{4}e^{-2E/\hbar\omega}
\left(\frac{2E}{\hbar\omega}\right)^{2+l+m+m'+L+M+M'}\\
\times \Gamma(3+l+m+m'+L+M+M')
\Gamma(-(3+l+m+m'+L+M+M'),-\frac{2E}{\hbar\omega})
\end{multline}
\begin{equation}\label{eqn:integral2}
\int_0^{\pi/2}d\phi \
(cos\phi)^{2(1+l+m+m')}(sin\phi)^{2(1+L+M+M')}=
\frac{\Gamma(l+m+m'+3/2)\Gamma(L+M+M'+3/2)}{2\Gamma(3+l+m+m'+L+M+M')}.
\end{equation}

This procedure generalizes for matrix elements of $\frac{1}{E-T}$ for
higher-body systems.  For an A-body problem within Jacobi coordinates,
a change of variables to (A-1)-dimensional spherical coordinates is
needed so that the resulting (A-1) integrals `factorize' (e.g. those
in square brackets of Eq.~\ref{eqn:3Mec}).
\end{widetext}

\begin{acknowledgments}
  We thank Andreas Nogga for his insightful discussions and
  contribution of codes for calculating the permutation operator
  $\Pi$.  This work was supported in part by the Division of Nuclear
  Physics, Office of Science, U.S. Department of Energy, and in part
  under the auspices of the U. S. Department of Energy by the
  University of California, Lawrence Livermore National Laboratory
  under contract No.  W-7405-Eng-48.
\end{acknowledgments}



\begin{table}
\caption{Dimension of included space for various sizes $\Lambda$ and isospin channel $T$ 
for the three-body system.\label{tab:dimension}}
\begin{center}
\begin{tabular}
{|c|c|c|c|}
\hline\hline
$\Lambda$ & \# of $T=1/2$ states & \# of $T=3/2$ states & Total  \\ \hline
 0 & 1 & 0 & 1 \\ \hline
4 & 15 & 7 & 22 \\ \hline
10 & 108 & 63 & 161 \\ \hline
20 & 632 & 314 & 946 \\ \hline
30 & 1906 & 950 & 2856 \\ \hline
40 & 4263 & 2128 & 6391 \\ \hline
50 & 8037 & 4014 & 12051 \\ \hline
\hline
\end{tabular}
\end{center}
\end{table}
\begin{table}
\caption[Triton B.E. at different oscillator $b$s]{Triton B.E. at
 different oscillator $b$s and $\Lambda$s.  These results ignore the
 small isospin-symmetry breaking effects included in the Av18
 potential.  Hence isospin is conserved (T=1/2).  Numbers in parentheses refer to the
 lowest eigenvalue of the bare $H$ within the corresponding model
 space. Results in MeV.}  
\begin{center}
\begin{tabular}
{|c|c|c|c|}
\hline\hline
$\Lambda$ & b=0.83 fm & b=1.17 fm & b=1.66 fm \\ \hline
0  & -7.6147 (218.533) & -7.6144 (74.113) & -7.6148 (26.480)   \\ \hline
2  & -7.6143 (65.260)  & -7.6144 (26.732) & -7.6145 (12.556)   \\ \hline
4  & -7.6144 (26.226)  & -7.6141 (15.840) & -7.6141 (9.068)   \\ \hline
6  & -7.6179 (13.268)  & -7.6140 (9.899)  & -7.6141 (6.524)   \\ \hline
8  & -7.6179 (5.985)   & -7.6136 (6.725)  & -7.6139 (5.257)   \\ \hline
10 & -7.6182 (1.606)   & -7.6131 (4.156)  & -7.6138 (4.093)   \\ \hline
\hline
\end{tabular}
\end{center}
\label{tab:triton_results}
\end{table}
\begin{table}
\caption{Triton B.E. at different oscillator $b$s and $\Lambda$s.
 These results include isospin-symmetry breaking parts of the Av18
 potential.  Hence the ground state consists of both T=1/2 \& 3/2
 channels. Results in MeV.}
\begin{center}
\begin{tabular}
{|c|c|c|c|}
\hline\hline
$\Lambda$ & b=0.83 fm & b=1.17 fm & b=1.66 fm \\ \hline
0  & -7.6219  & -7.6217 & -7.6213    \\ \hline
2  & -7.6230    & -7.6209  & -7.6209    \\ \hline
4  & -7.6255    & -7.6217   & -7.6206    \\ \hline
\hline
\end{tabular}
\end{center}
\label{tab:tritonNP3half}
\end{table}
\begin{table}
\caption[Triton $<0\hbar\omega|H_{eff}|0\hbar\omega>$ matrix elements 
at different oscillator $b$s and model sizes $\Lambda$]{Triton 
$<0\hbar\omega|H_{eff}|0\hbar\omega>$ matrix element at different 
oscillator $b$s and model sizes $\Lambda$.  Within each cell, the 
top number represents $<T_{eff}>$, the middle number \\
$3<V_{12,eff}^{(2+1)}>$, and the bottom number $3<V_{12,eff}^{(3+0)}>$.  Numbers are in MeV.}
\begin{center}
\begin{tabular}
{|c||c|c|c|c|c|c||c|}
\hline\hline
$\hbar\omega\ \Lambda$ & 0 & 2 & 4 & 6 & 8 & 10 & $\infty$ \\ \hline\hline 
60 & 64.51  & 90.00  & 90.00  & 90.00  & 90.00  & 90.00  & 90.00  \\
   & -68.36 & -38.41 & -32.54 & -26.40 & -21.03 & -16.52 & 128.50 \\ 
   & -3.76  & 8.26   & 22.38  & 37.25  & 51.34  & 64.06  & 0.00   \\ \hline
30 & 33.59  & 45.00  & 45.00  & 45.00  & 45.00  & 45.00  & 45.00  \\
   & -45.11 & -47.20 & -45.61 & -44.05 & -42.50 & -40.99 & 29.10  \\ 
   & 3.91   & 3.71   & 6.17   & 8.96   & 11.96  & 15.03  & 0.00   \\ \hline
15 & 17.67  & 22.50  & 22.50  & 22.50  & 22.50  & 22.50  & 22.50  \\
   & -26.37 & -31.54 & -30.43 & -29.73 & -29.16 & -28.66 & 3.98   \\ 
   & 1.08   & 2.43   & 4.16   & 5.03   & 5.68   & 6.26   & 0.00   \\ \hline
\hline
\end{tabular}
\end{center}
\label{tab:00tritonME}
\end{table}
\begin{table}
\caption[Triton $<0\hbar\omega|H_{eff}|2\hbar\omega>$ matrix elements 
at different oscillator $b$s and model sizes $\Lambda$]{A particular 
triton $<0\hbar\omega|H_{eff}|2\hbar\omega>$ matrix element at
different oscillator $b$s and model sizes $\Lambda$.  Within each 
cell, the top number represents $3<V_{12,eff}^{(2+1)}>$, and the 
bottom number $3<V_{12,eff}^{(3+0)}>$.  Numbers are in MeV.  Note that $T_{eff}=0$ for this case.}
\begin{center}
\begin{tabular}
{|c||c|c|c|c|c||c|}
\hline\hline
$\hbar\omega\ \Lambda$ & 2 & 4 & 6 & 8 & 10 & $\infty$ \\ \hline\hline 
60 & -33.149 & -47.107 & -46.936 & -46.873 & -46.837 & -45.375 \\ 
   & 0.146   & -1.771  & -1.692  & -1.016  & -0.330  & 0.000   \\ \hline
30 & -13.400 & -18.898 & -18.772 & -18.665 & -18.564 & -18.952 \\ 
   & 0.641   & -0.197  & -0.655  & -0.920  & -1.065  & 0.000   \\ \hline
15 & -5.053  & -6.964  & -6.911  & -6.884  & -6.857  & -6.769  \\ 
   & 0.398   & 0.385   & 0.296   & 0.194   & 0.104   & 0.000   \\ \hline
\hline
\end{tabular}
\end{center}
\label{tab:02tritonME}
\end{table}
\begin{table}
\caption[Triton $<2\hbar\omega|H_{eff}|2\hbar\omega>$ matrix elements 
at different oscillator $b$s and model sizes $\Lambda$]{A particular 
Triton $<2\hbar\omega|H_{eff}|2\hbar\omega>$ matrix element at
different oscillator $b$s and model sizes $\Lambda$.  Within each 
cell, the top number represents $<T_{eff}>$, the middle number 
$3<V_{12,eff}^{(2+1)}>$, and the bottom number $3<V_{12,eff}^{(3+0)}>$.  Numbers are in MeV.}
\begin{center}
\begin{tabular}
{|c||c|c|c|c|c||c|}
\hline\hline
$\hbar\omega\ \Lambda$ & 2 & 4 & 6 & 8 & 10 & $\infty$ \\ \hline\hline 
60 & 122.210  & 150.000  & 150.000  & 150.000  & 150.000  & 150.000 \\
   & -11.616  & 11.714   & 15.141   & 18.365   & 20.788   & 89.884 \\ 
   & 10.381   & 18.639   & 25.302   & 31.312   & 36.794   & 0.000   \\ \hline
30 & 61.986   & 75.000   & 75.000   & 75.000   & 75.000   & 75.000  \\
   & -16.159  & -12.223  & -11.338  & -10.376  & -9.496   & 21.034  \\ 
   & 4.571    & 7.211    & 8.547    & 9.790    & 11.033   & 0.000   \\ \hline
15 & 31.679   & 37.500   & 37.500   & 37.500   & 37.500   & 37.500  \\
   & -10.649  & -10.579  & -10.162  & -9.805   & -9.509   & 4.019  \\ 
   & 2.144    & 3.365    & 3.782    & 4.071    & 4.335    & 0.000   \\ \hline
\hline
\end{tabular}
\end{center}
\label{tab:22tritonME}
\end{table}
\begin{table}
\caption{Values of $b$ where the contribution from
  $V^{(3+0)}_{12,eff}$ vanishes for a few included spaces sizes.  Note
  that for each included space, there are two values.  }
\begin{center}
\begin{tabular}
{|c||c|c|c|}
\hline\hline
$\Lambda$ & 0 & 2 & 4 \\ \hline\hline 
b (fm) & 0.895, 1.838  & 0.562, 2.219 & 0.375, 2.706 \\
\hline
\hline
\end{tabular}
\end{center}
\label{tab:b_zeros}
\end{table}

\end{document}